\documentclass[letterpaper,12pt]{JHEP}
\usepackage{amsmath,amssymb}
\usepackage{epsfig}
\usepackage{cite}
\usepackage[english]{babel}

\newcommand{\dual}{{}^*}

\newcommand{\dd}{{\rm d}}

\newcommand{\re}{\mathrm{Re}}
\newcommand{\im}{\mathrm{Im}}

\newcommand{\eq}{\begin{equation}}
\newcommand{\feq}{\end{equation}}
\newcommand{\eqn}{\begin{eqnarray}}
\newcommand{\feqn}{\end{eqnarray}}
\newcommand{\arr}{\begin{eqnarray*}}
\newcommand{\farr}{\end{eqnarray*}}

\newcommand{\RR}{{\cal R}}
\newcommand{\DD}{{\cal D}}

\newcommand{\F}{{\cal F}}
\newcommand{\G}{{\cal G}}
\newcommand{\p}{\partial}

\font\mybb=msbm10 at 12pt
\def\bb#1{\hbox{\mybb#1}}
\def\bZ {\bb{Z}}
\def\bR {\bb{R}}

\newcommand{\HH}{{\mathbb{H}}}

\def\al{\alpha}

\def\ga{\gamma}

\def\ep{\epsilon}
\def\eps{\varepsilon}

\def\si{\sigma}
\def\om{\omega}

\def\Ga{\Gamma}

\title{More on BPS solutions of $N=2$, $D=4$ gauged supergravity}
\author{Sergio L.~Cacciatori \\
Dipartimento di Matematica dell'Universit\`a di Milano \\
Via Saldini 50, I-20133 Milano and \\
INFN, Sezione di Milano, Via Celoria 16, I-20133 Milano. \\
E-mail: \email{cacciatori@mi.infn.it}}
\author{Marco M.Caldarelli, Dietmar Klemm and Diego S.~Mansi \\
Dipartimento di Fisica dell'Universit\`a di Milano \\
Via Celoria 16, I-20133 Milano and \\
INFN, Sezione di Milano, Via Celoria 16, I-20133 Milano. \\
E-mail: \email{marco.caldarelli@mi.infn.it,
               dietmar.klemm@mi.infn.it,
               diego.mansi@mi.infn.it}}
\preprint{IFUM-796-FT \\
hep-th/0406238}

\abstract{We deepen and refine the classification of supersymmetric solutions
to $N=2$, $D=4$ gauged supergravity obtained in a previous paper. In the case
where the Killing vector constructed from the Killing spinor is timelike,
it is shown that the nonlinear partial differential equations determining the
BPS solutions can be derived from a variational principle. The corresponding
action enjoys a solution-generating PSL$(2,\bR)$ symmetry. In certain subcases
the system reduces to different known theories, like two-dimensional dilaton
gravity or the dimensionally reduced gravitational Chern-Simons theory. We find
new supersymmetric solutions including, among others, kinks that
interpolate between two AdS$_4$ vacua, electrovac waves on anti-Nariai
spacetimes, or generalized Robinson-Trautman solutions. In the case
where the Killing vector is null, we obtain a complete
classification. The one quarter and one half supersymmetric solutions
are determined explicitely, and it is shown that the fraction of three
quarters of supersymmetry cannot be preserved. Finally, the general
lightlike configuration is uplifted to eleven-dimensional
supergravity.
}


\begin{document}

\section{Introduction}
\label{intro}

Supersymmetric solutions have played an important role in recent progress
in string theory. This makes it desirable to obtain a systematic classification
of BPS solutions to various supergravity theories. Apart from the seminal work
by Tod \cite{Tod:pm}, who wrote down all metrics admitting supercovariantly
constant spinors in $N=2$, $D=4$ ungauged supergravity, progress in
this direction has been made mainly during the last two years using
the mathematical concept of G-structures \cite{Gauntlett:2002sc}. The
basic strategy is to assume the existence of at least one Killing
spinor, and to construct differential forms as bilinears from this
supercovariantly constant spinor. These forms obey several algebraic
and differential equations that can be used to deduce the metric and
the other bosonic supergravity fields. This formalism has been
successfully applied to several supergravity theories in diverse
dimensions \cite{Gauntlett:2002nw,Caldarelli:2003pb}.
A common feature is that the Killing vector constructed from the
Killing spinor is either timelike or lightlike, so that the solutions
fall into two (partially overlapping) classes.

In this paper we will deepen and refine the classification
of supersymmetric solutions of $N=2$, $D=4$ gauged supergravity obtained
in \cite{Caldarelli:2003pb}\footnote{For the inclusion of sources
see \cite{Caldarelli:2003wh}.}. The motivation for this is threefold: first of
all, as pointed out above, having at hand a systematic approach to construct
BPS solutions avoids the use of special ans\"atze that cover only a certain
subclass of supersymmetric configurations. The second reason comes from the
AdS$_4$/CFT$_3$ correspondence. In three dimensions there is a rich web of
conformal field theories that have many interesting and physically relevant
perturbative as well as non-perturbative properties. It might turn out that
some of these CFTs have an AdS$_4$ dual. Then, supergravity vacua with
less than maximal supersymmetry may have an interpretation on the CFT side
as an expansion of the theory around non-zero vacuum expectation values of
certain operators. Thirdly, Mathur et al.~proposed recently a gravity picture
of black hole microstates \cite{Mathur:2003hj}.
According to \cite{Mathur:2003hj}, these microstates are completely
regular (i.~e.~, horizonless and nonsingular) supergravity solutions
carrying the same charges and mass as the black hole. Coarse graining
over the microstates (in a sense that is explained in
\cite{Mathur:2003hj}) yields the Bekenstein-Hawking entropy of the
black hole. In the case of the D1-D5-P black hole, a set of such
solutions has been constructed in
\cite{Mathur:2003hj,Lunin:2004uu,Giusto:2004id}.
It is plausible that, if the black hole under consideration is
supersymmetric, then also the gravity microstates should preserve the
same amount of supersymmetry. Thus, a classification of BPS
solutions of $N=2$, $D=4$ gauged supergravity should be relevant for
the construction of microstates for supersymmetric AdS black holes.

The remainder of this paper is organized as follows. In section
\ref{timelikecase}, we discuss the case where the Killing vector
$V^{\mu} = i\bar{\epsilon}\Gamma^{\mu}\epsilon$ obtained from the Killing
spinor $\epsilon$ is timelike. The general form of the metric and the
electromagnetic field strength was given in \cite{Caldarelli:2003pb}.
The geometry is characterized by some functions that satisfy highly nonlinear
partial differential equations. Here, we shall reveal some of the mathematical
structure behind these equations, and present many new solutions,
which give rise to new supersymmetric supergravity configurations.
In particular, we show that the differential equations follow from
an action principle in three dimensions. This action enjoys a
PSL$(2,\bR)$ invariance, which can be used to generate new solutions from
known ones. Quite surprisingly, it turns out that a certain subclass
of solutions to the general equations is governed by the dimensionally
reduced gravitational Chern-Simons action. The deeper significance of this
intriguing fact is rather obscure and deserves further investigations.
Among the new solutions that we shall present there are, among others,
deformations of AdS$_4$, kink solutions that interpolate between two
maximally supersymmetric AdS vacua, and charged supersymmetric generalizations
of the Robinson-Trautman type geometries.\\
In section \ref{lightlike} we give a complete classification of the lightlike
case, where the general supersymmetric solution is given by an electrovac
AdS travelling wave \cite{Caldarelli:2003pb}, whose profile satisfies
a generalized Siklos equation. It is shown that a configuration
which admits a null Killing spinor, i.~e.~, a Killing spinor which can be
used to construct a null Killing vector, is either one quarter or one
half supersymmetric. The fraction three quarters of supersymmetry cannot
be preserved in this case. For vanishing electromagnetic fields, the solution
is always one quarter supersymmetric.
The explicit form of the wave profile for
the one half BPS case is given. We shall furthermore show that for
half-supersymmetric solutions, the second Killing spinor gives rise to
a timelike Killing vector, which implies that these waves are also
solutions of the timelike case. Finally, the general lightlike geometry
is lifted to a solution of eleven-dimensional supergravity.\\
We conclude in section \ref{finalrem} with some final remarks.
The appendices contain our conventions and notations as well as some
supplementary material.

%%%%%%%%%%%%%%%%%%%%%%%%%%%%%%%%%%%%%%%%%%%%%%%%%%%%%%%%%%%%%%%%%%%%%%%%%%%
\section{The timelike case}\label{timelikecase}

Let us briefly recall the results of \cite{Caldarelli:2003pb} for timelike
$V^{\mu}$ (rewritten here in a slightly more compact form). The general
BPS solution reads\footnote{We have chosen the conformal gauge for the
  two-metric $h_{ij}$ appearing in \cite{Caldarelli:2003pb}.}
\begin{eqnarray}
\dd s^2 &=& -\frac 4{\ell^2 F\bar F}\left(\dd t + \omega_i \dd x^i\right)^2 + \frac{\ell^2 F\bar F}4
       \left[\dd z^2 + e^{2\phi}\left(\dd x^2 + \dd y^2\right)\right]\,, \label{metric} \\
{\cal F} &=& \frac{\ell^2}4 F\bar F\left[V\wedge\dd f+\dual\left(V\wedge\left(\dd
             g+\frac1\ell \dd z\right)\right)\right]\,, \nonumber
\end{eqnarray}
where $i=1,2$; $x^1 = x$, $x^2=y$, and we defined $\ell F=2i/(f-ig)$,
with $f=\bar{\epsilon}\epsilon$ and $g=i\bar{\epsilon}\Gamma_5\epsilon$.
Here, $1/\ell$ is the minimal coupling between the graviphoton and the
gravitini, which is related to the cosmological constant by
$\Lambda = -3\ell^{-2}$, and ${\cal F}$ denotes the electromagnetic field strength.
The timelike Killing vector is given by $V = \partial_t$.
The functions $\phi, F, \bar F$, that depend on $x,y,z$, are determined by
the system
\begin{eqnarray}
\Delta F + e^{2\phi}[F^3 + 3FF' + F''] &=& 0\,, \label{F} \\
\Delta \phi + \frac 12 e^{2\phi}[F' + \bar{F}' + F^2 + \bar{F}^2 - F\bar F] &=& 0\,,
\label{phi} \\
\phi' - \re (F) &=& 0\,, \label{ReF}
\end{eqnarray}
where $\Delta = \partial^2_x + \partial^2_y$, and a prime denotes differentiation
with respect to $z$. (\ref{F}) comes from the combined Maxwell equation and Bianchi
identity, whereas (\ref{phi}) results from the integrability condition for the
Killing spinor $\epsilon$. Finally, the shift vector $\omega$ is obtained
from\footnote{It can be shown that the integrability conditions for (\ref{omega})
follow from the Maxwell equations.}
\begin{eqnarray}
\partial_z \omega_i &=& \frac{\ell^4}8 (F\bar F)^2 \epsilon_{ij}(f\partial_j g
- g\partial_j f)\,, \nonumber \\
\partial_i \omega_j - \partial_j \omega_i &=& \frac{\ell^4}8 (F\bar F)^2 e^{2\phi}
\epsilon_{ij}\left(f\partial_z g - g\partial_z f + \frac{2f}{\ell}\right)\,, \label{omega}
\end{eqnarray}
with $\epsilon_{12} = 1$.

\subsection{Symmetries and properties of the equations}

Before presenting new solutions of the timelike case, let us study some
general properties of the system (\ref{F}) -- (\ref{ReF}).
First of all, we note that it is invariant under PSL$(2,\bR)$ transformations
\begin{equation}
z \to \frac{az + b}{cz + d}\,, \qquad ad - bc = 1\,, \label{PSLz}
\end{equation}
if the fields transform according to
\begin{eqnarray}
F &\to & (cz + d)^2 F -\partial_z (cz + d)^2\,, \nonumber \\
\phi &\to & \phi - 2\ln(cz + d)\,. \label{PSLphi}
\end{eqnarray}
This means that $F$ has a connection-like transformation behaviour,
whereas $\phi$ transforms like a Liouville field. In appendix
\ref{appPSL2R} we show, using the supersymmetric
Reissner-Nordstr\"om-Taub-NUT-AdS$_4$ geometry as an example, that this
symmetry is nontrivial, i.~e.~, it can be used to generate new
solutions.

If we introduce the complex coordinates $\zeta = x + iy$, $\bar\zeta = x - iy$,
we see that the eqns.~(\ref{F}) -- (\ref{ReF}) enjoy an additional
infinite-dimensional conformal symmetry
\begin{equation}
\zeta \to w(\zeta)\,, \qquad \bar\zeta \to \bar w(\bar\zeta)\,, \qquad F \to F\,,
\qquad \phi \to \phi - \frac 12 \ln\left(\frac{dw}{d\zeta}\frac{d\bar w}{d\bar\zeta}
\right)\,,
\end{equation}
where $w(\zeta)$ and $\bar w(\bar\zeta)$ denote arbitrary holomorphic and
antiholomorphic functions respectively. However, it is easy to
see that, from the four-dimensional point of view, this represents merely a
diffeomorphism that preserves the conformal gauge for the two-metric
$e^{2\phi}(dx^2 + dy^2)$. Thus, unlike the PSL$(2,\bR)$
transformations above, this symmetry cannot be used to generate new
solutions from known ones.

Decomposing $F$ into its real and imaginary part, $F = A + iB$, we see that the
real part of eqn.~(\ref{F}) follows from (\ref{phi}) and (\ref{ReF}),
so that the remaining system is
\begin{eqnarray}
\Delta B + e^{2\phi}\left[3\phi^{\prime 2} B - B^3 + 3\phi' B' + 3B\phi'' + B''\right] &=& 0\,, \label{B} \\
\Delta \phi + \frac 12 e^{2\phi}\left[2\phi'' + {\phi'}^2 - 3B^2\right] &=& 0\,, \label{phinew}
\end{eqnarray}
together with $A = \phi'$.
The equations (\ref{B}), (\ref{phinew}) can be derived from the action
\begin{eqnarray}
S = \int \dd^2x\,\dd z\,\left[\nabla B\cdot\nabla \phi + \frac 12 e^{2\phi}
\left(B^3 + 2B'\phi' + 3B{\phi'}^2\right)\right]\,.
    \label{actionBphi}
\end{eqnarray}
This action is also invariant under the above PSL$(2,\bR)$
transformations, with $B$ transforming like the imaginary part of $F$,
i.~e.~,
\begin{equation}
B \to (cz + d)^2 B\,.
\end{equation}
$B$ transforms thus like a conformal field of weight two.

Multiplying eqn.~(\ref{F}) by $\bar F$ and subtracting the complex conjugate
yields the conservation law
\begin{equation}
\partial_i j_i + \rho' = 0\,,
\end{equation}
with the current $j_i$ and the ``charge density'' $\rho$ given
respectively by
\begin{eqnarray}
  j_i &=& \frac 1{2i}\left(\bar F \partial_i F - F \partial_i \bar F\right)
  = \phi' \partial_i B - B \partial_i \phi'\,, \nonumber \\
  \rho &=& \frac 1{2i}\left(\bar \lambda'\,\lambda'' - \lambda'\,\bar \lambda''\right)
  = e^{2\phi}\left[\left({\phi'}^2 + B^2\right)B + \phi' B' - B\phi''\right]\,,
\end{eqnarray}
where $\lambda = \exp\int F \dd z$. This current conservation is
presumably related to the PSL$(2,\bR)$ invariance of the action
(\ref{actionBphi}), although we did not check this explicitely.

In the ``purely magnetic'' case ($f=0$), one has $B=0$, so that the
only equation to solve is
\begin{equation}
\Delta \phi + \frac 12 e^{2\phi}\left[{\phi'}^2 + 2\phi''\right] = 0\,.
\end{equation}
This is similar to the continuous $(SU(\infty))$ Toda equation \cite{Boyer:mm}
(or heavenly equation)
\begin{equation}
\Delta \phi + \frac 12 \partial^2_z e^{2\phi} = 0\,,
\end{equation}
which determines self-dual Einstein metrics that admit at least one rotational
Killing vector \cite{Boyer:mm}.

In the ``purely electric'' case ($g=0$) we have $A=0$, and thus $\phi$
is independent of $z$.
(\ref{phinew}) implies then that also $B$ does not depend on $z$, and
the equations (\ref{B}), (\ref{phinew}) reduce to
\begin{eqnarray}
\Delta B - e^{2\phi}B^3 &=& 0\,, \label{Bg=0} \\
\Delta \phi - \frac 32 e^{2\phi}B^2 &=& 0\,. \label{phig=0}
\end{eqnarray}
These equations follow from the two--dimensional dilaton gravity action
\begin{equation}
S = \int \dd^2 x \sqrt g \left[BR + B^3\right]\,, \label{dilgrav}
\end{equation}
if we use the conformal gauge $g_{ij}\dd x^i \dd x^j = e^{2\phi}(\dd x^2 + \dd y^2)$.
However, the equations of motion following from (\ref{dilgrav}) contain
also the constraints $\delta S/\delta g^{ij} = 0$ (whose trace yields
(\ref{Bg=0})), and therefore they are more restrictive than the system
(\ref{Bg=0}), (\ref{phig=0}). Of course, every solution that extremizes
the action (\ref{dilgrav}) is a solution of our system, but not vice versa.
It is interesting to note that (\ref{dilgrav}) is similar to the action
that arises from Kaluza-Klein reduction of the three-dimensional gravitational
Chern-Simons term \cite{Guralnik:2003we}, with the only difference that here
$B$ is a fundamental field, whereas the field arising in \cite{Guralnik:2003we}
is the curl of a vector potential.

We can actually obtain exactly the model considered in \cite{Guralnik:2003we}
by allowing for $z$-dependent $A$ instead of setting $A=0$,
i.~e.~, we set $F=A(z) + iB(x,y,z)$. From (\ref{ReF}) one obtains then
\begin{equation}
\phi = \int A(z)\,\dd z + \gamma(x,y)\,,
\end{equation}
with $\gamma(x,y)$ denoting a function of $x,y$ alone. If we define
\begin{equation}
  \beta(x,y,z) \equiv B\exp\int A\,dz\,,
\label{defbeta}\end{equation}
eqn.~(\ref{phi}) implies that
$\beta^2$ separates into a part that depends only on $z$ and a function
of $x,y$,
\begin{displaymath}
\beta^2(x,y,z) = Q(z) + R(x,y)\,.
\end{displaymath}
Let us consider the case $Q = 0$. (\ref{phi}) yields then
\begin{eqnarray}
\Delta \gamma + \frac 12 e^{2\gamma}\left(k - 3\beta^2\right) &=& 0\,, \label{gamma} \\
e^{2\int A\,dz}\left(2A' + A^2\right) &=& k\,, \nonumber
\end{eqnarray}
where $k$ denotes an arbitrary constant. The latter equation is solved by
\begin{equation}
A = \frac{2az + b}{az^2 + bz + c}\,, \label{A}
\end{equation}
with $a,b,c$ being real integration constants obeying $4ac - b^2 = k$.
Finally, from (\ref{B}) one obtains
\begin{equation}
\Delta \beta + e^{2\gamma}\left(k\beta - \beta^3\right) = 0\,. \label{beta}
\end{equation}
We see that the equations (\ref{gamma}), (\ref{beta}) represent
generalizations of (\ref{phig=0}) and (\ref{Bg=0}) respectively, which
are recovered in the case $k=0$\footnote{Actually also for $k=0$ $A$ can still
depend on $z$, but in this case $A(z)$ is related to $A=0$ by a PSL$(2,\bR)$
transformation (\ref{PSLz}), (\ref{PSLphi}). To see this, one observes that
for $k=0$ one has $2A' + A^2 = 0$ and thus $A = 2/(z + \tilde b)$, where
$\tilde b$ denotes an arbitrary integration constant. Now one performs a
PSL$(2,\bR)$ transformation with $d = c\tilde b$. The transformed $A$ will
then vanish. Note that it is no more possible to transform $A$ to zero if
$k \neq 0$.}.
What comes as a surprise is that
(\ref{gamma}) and (\ref{beta}) follow from the dimensionally reduced
gravitational Chern-Simons action
\begin{equation}
S = \int \dd^2 x \sqrt g \left[\beta R + \beta^3\right]\,, \label{CS}
\end{equation}
if we use the conformal gauge $g_{ij}\dd x^i \dd x^j = e^{2\gamma}(\dd x^2 + \dd y^2)$.
Note that in (\ref{CS}) $\beta$ is not a fundamental field, rather it is
the curl of a vector potential, $\sqrt g \epsilon_{ij}\beta = \partial_i A_j
- \partial_j A_i$. Did we vary $\beta$ instead of $A_i$ in the action, we
would obtain the equations (\ref{Bg=0}), (\ref{phig=0}), i.~e.~, the case
$k=0$. Like before, the equations of motion following from the
action (\ref{CS}) are slightly stronger than our system, which does not
include the traceless part of the constraints $\delta S/\delta g^{ij} = 0$.

We will now present some solutions of the system (\ref{F}) -- (\ref{ReF}),
which give rise to new BPS states of $N=2$, $D=4$ gauged supergravity.

%%%%%%%%%%%%%%%%%%%%%%%%%%%%%%%%%%%%%%%%%%%%%%%%%%%%%%%%%%%%%%%%%%%%%%%%%%%%%
\subsection{``Purely electric'' solutions ($g=0$)}
\label{electric}

First, we consider the case $g=0$ and assume that $B$ and $\phi$ depend only
on $x$. In the coordinates system (\ref{metric}), the $g=0$
condition corresponds to have a vanishing magnetic field, and
therefore we shall refer to these solutions as the ``purely electric''
ones. If we use the ansatz $B = x^{\alpha}$, equations
(\ref{Bg=0}) and (\ref{phig=0}) are satisfied for $\alpha = 2$ or
$\alpha = -1/3$. The former value of $\alpha$ yields the Petrov type I
solution obtained in \cite{Caldarelli:2003pb}, whereas for the latter one
we get (after solving eqn.~(\ref{omega}) for $\omega$) the four-dimensional geometry
\begin{equation}
\dd s^2 = -\frac{4x^{2/3}}{\ell^2}\left(\dd t - \frac{\ell^2}{6x^{4/3}}\dd y\right)^2
       + \frac{\ell^2}{4x^{2/3}}\dd z^2 + \frac{\ell^2}{9x^2}(\dd x^2 + \dd y^2)\,,
\end{equation}
and the electromagnetic field strength
\begin{equation}
{\cal F} = -\frac{2}{3\ell x^{2/3}}\dd t \wedge \dd x\,.
\end{equation}
Introducing the new coordinates
\begin{displaymath}
r = 2x^{1/3}\,, \qquad u = t\,, \qquad v = \frac{8}{3\ell^2}y\,,
\end{displaymath}
the solution can be written in the form
\begin{eqnarray}
\dd s^2 &=& \frac{\ell^2}{r^2}\left[-\frac{r^4}{\ell^4}\dd u^2 + 2\dd u\dd v + \dd r^2 + \dd z^2\right]\,,
         \label{alpha=-1/3} \\
{\cal F} &=& -\frac 1{\ell} \dd u \wedge \dd r\,. \nonumber
\end{eqnarray}
This represents an electrovac AdS travelling wave; in section
\ref{lightlike} we will show that it is also a solution of the
lightlike case and that it has two Killing spinors, one that gives the
timelike Killing vector $V = \partial_u$ and another that yields the
lightlike Killing vector $U = \partial_v$. In addition to $U$ and $V$,
the geometry (\ref{alpha=-1/3}) admits the Killing vectors
\begin{eqnarray}
Z &=& \partial_z\,, \qquad K = u\partial_z - z\partial_v\,, \nonumber \\
D &=& u\partial_u - z\partial_z - r\partial_r - 3v\partial_v\,,
\end{eqnarray}
which obey the commutation relations
\begin{eqnarray}
[D,V] &=& -V\,, \qquad [D,Z] = Z\,, \qquad [D,U] = 3U\,, \nonumber \\
{[}K,V] &=& -Z\,, \qquad [K,Z] = U\,, \qquad [D,K] = 2K\,. \nonumber
\end{eqnarray}
The isometry group acts transitively on the spacetime, which is thus
homogeneous. A computation of the Weyl scalars shows that its Petrov
type is N.

%%%%%%%%%%%%%%%%%%%%%%%%%%%%%%%%%%%%%%%%%%%%%%%%%%%%%%%%%%%%%%%%%%%%%%%%%%%%%
\subsection{``Purely magnetic'' solutions ($f=0$)}
\label{magnetic}

When $f=0$, the function $F=A$ is real, and the equations reduce to the
system (\ref{gamma}), (\ref{A}) and (\ref{beta}),
with $\beta=0$ (which is a simple solution of eqn.~(\ref{beta})).
These are ``purely magnetic'' solutions, in the sense that in these
coordinates the electric field vanishes, and are easily seen to be
also static. Eqn.~(\ref{gamma}) becomes then the Liouville equation
\begin{equation}
  \Delta\gamma + \frac k2 e^{2\gamma} = 0\,, \label{Liouville}
\end{equation}
which describes the metrics on euclidean 2-manifolds with constant
curvature $k$.

Let us consider first the case $a=0$ in (\ref{A}). Without loss of
generality we set $b=1$, $c=0$, so that $k = -1$ and $A = 1/z$.
As a solution of the Liouville equation we choose $e^{2\gamma} = 2/x^2$.
This yields the Bertotti-Robinson type AdS$_2\times\HH^2$ spacetime,
with a purely magnetic Maxwell field,
\begin{eqnarray}
    \dd s^2 &=& -\frac{4z^2}{\ell^2}\dd t^2+\frac{\ell^2}{4z^2}
             \dd z^2 + \frac{\ell^2}{2x^2}(\dd x^2 + \dd y^2), \\
    {\cal F} &=& -\frac{\ell}{2x^2} \dd x \wedge \dd y\,.\nonumber
\end{eqnarray}

In \cite{Cacciatori:1999rp}, this configuration was shown to preserve
half of the supersymmetry, and to admit an
$osp(2|2) \oplus so(2,1) \cong su(1,1|1) \oplus su(1,1)$
isometry superalgebra.

For $a$ different from zero we can set without loss of generality
$a = 1/\ell$, $b = 0$, $c = k\ell/4$, so that
\begin{equation}
A = \frac{2z}{z^2 + \frac{k\ell^2}4}\,.
\end{equation}

If $k=0$ we have $A = 2/z$ and $\gamma(x,y)$ is harmonic, so that the 2-manifold
with metric $e^{2\gamma}(dx^2 + dy^2)$ is flat. The choice $\gamma = 0$ leads
to the maximally supersymmetric AdS$_4$ vacuum solution,
\begin{eqnarray}
    ds^2 &=& -\frac{z^2}{\ell^2} \dd t^2 + \frac{\ell^2}{z^2}
             \dd z^2 + z^2 (\dd x^2 + \dd y^2)\,, \\
    {\cal F} &=& 0\,.
\end{eqnarray}
As was shown in \cite{Caldarelli:2003pb}, this is also the only configuration
with maximal supersymmetry.

For $k < 0$ the solutions of the Liouville equation (\ref{Liouville}) are
classified according to their monodromy (cf.~e.~g.~\cite{Seiberg:1990eb}).
If we introduce polar coordinates $(r, \sigma)$ on the $(x,y)$-plane,
we have the following $\sigma$-independent solutions \cite{Seiberg:1990eb}:
\begin{itemize}
\item \textbf{Elliptic monodromy:}
\begin{equation}
e^{2\gamma} = -\frac{2m^2}{kr^2 \sinh^2(m\ln r)}\,,
\end{equation}
\item \textbf{Parabolic monodromy:}
\begin{equation}
e^{2\gamma} = -\frac{2}{kr^2 \ln^2 r}\,,
\end{equation}
\item \textbf{Hyperbolic monodromy:}
\begin{equation}
e^{2\gamma} = -\frac{2m^2}{kr^2 \sin^2(m\ln r)}\,.
\end{equation}
\end{itemize}
Here, $m$ is a constant related to the Liouville-momentum.
The corresponding supergravity solution is then given by
\begin{eqnarray}
\dd s^2 &=& \left(\frac{z}{\ell} + \frac{k\ell}{4z}\right)^2 \dd t^2 +
         \frac{\dd z^2}{\left(\frac{z}{\ell} + \frac{k\ell}{4z}\right)^2} +
         z^2 e^{2\gamma}(\dd x^2 + \dd y^2)\,, \label{magnmonop} \\
{\cal F} &=& \frac{k\ell}4 e^{2\gamma} \dd x \wedge \dd y\,. \nonumber
\end{eqnarray}
In the case of elliptic monodromy and $m=1$, this reduces to the configuration
found in \cite{Caldarelli:1998hg}\footnote{For generalizations to five dimensions
see \cite{Chamseddine:1999xk,Klemm:2000nj}.}, which preserves one quarter of the
supersymmetry and admits an $s(2) \oplus su(1,1)$ isometry
superalgebra \cite{Cacciatori:1999rp}\footnote{$s(2)$ denotes the
  superalgebra introduced by E.~Witten to formulate supersymmetric
  quantum mechanics \cite{Witten:nf}.}.
(\ref{magnmonop}) represents an extremal black hole with event horizon at
$z = \sqrt{-k}\ell/2$.

For $k>0$ the four-dimensional metric and gauge field are again given by
(\ref{magnmonop}), with the Liouville field
\begin{equation}
e^{2\gamma} = \frac{2m^2}{kr^2 \cosh^2(m\ln r)}\,.
\end{equation}
Introducing the new coordinates $\theta, \varphi$ by $r^m = \tan\theta/2$,
$m\sigma = \varphi$, we get
\begin{equation}
e^{2\gamma}(\dd r^2 + r^2 \dd\sigma^2) = \frac 2k (\dd\theta^2 + \sin^2\theta \dd\varphi^2)\,.
\end{equation}
As $\sigma$ is identified modulo $2\pi$, we see that $\varphi \sim \varphi + 2\pi m$,
so that the effect of the parameter $m$ is to introduce a conical singularity ($0 < m < 1$)
or an excess angle ($m > 1$) on the north- and south pole of the two-sphere.
For $m=1$ there is no singularity, and the solution reduces to the one-quarter
supersymmetric magnetic monopole found by Romans \cite{Romans:1991nq}.
The magnetic charge of the solution reads
\begin{equation}
Q_m = \frac{1}{4\pi}\int {\cal F} = \frac{m\ell}2\,.
\end{equation}
If $m \neq 1$, there is a magnetic fluxline that passes through $x = y =0$,
which causes the magnetic charge to be different from the value $Q_m = \ell/2$
of Romans' solution.

%%%%%%%%%%%%%%%%%%%%%%%%%%%%%%%%%%%%%%%%%%%%%%%%%%%%%%%%%%%%%%%%%%%%%%%%%%%%%
\subsection{Solutions with $F=A(z)+iB(x,y,z)$.}

If we allow only for a $z$-dependence for $A$, then the conformal
factor $\phi$ of the transverse two-metric is given by (\ref{phi}),
where the function $\gamma(x,y)$ is an integration constant.
In two cases the equations can be further simplified, depending on the
form of the function $\beta$ defined in (\ref{defbeta}). The first
case is when $\beta$ depends only on $(x,y)$, and the second when
$\beta=\beta(z)$.

%____________________________________________________________________________%
\subsubsection{The case $\beta=\beta(x,y)$}

The simplest case of vanishing $\beta$ has been analyzed in
section~\ref{magnetic}.
By relaxing this condition and allowing for an $(x,y)$-dependence in
$\beta$, the BPS configurations are described by the system (\ref{gamma}),
(\ref{A}) and (\ref{beta}). Interesting solutions of these equations
can be found if we choose $a=1$, $b = c = 0$
in (\ref{A}), so that $k = 0$ and $A = 2/z$. Eqns.~(\ref{gamma}) and
(\ref{beta}) reduce to
\begin{eqnarray}
\Delta \gamma - \frac 32 e^{2\gamma} \beta^2 &=& 0\,,\\
\Delta \beta - e^{2\gamma}\beta^3 &=& 0\,.
\end{eqnarray}
The form of $A$ reminds us of AdS$_4$, which can be
recovered setting $\beta=0$. For this reason the spacetimes below can
be considered as modifications of AdS$_4$.
The important point is that these solutions are governed by the same set of
equations which describes ``purely electric'' solutions of
section~\ref{electric}, as can be checked by comparing the equations
above with (\ref{phig=0}) and (\ref{Bg=0}).
In section~\ref{electric} we found two solutions: $B=x^2$ and
$B=x^{-1/3}$ which yield spacetimes of Petrov type I and N
respectively.
One can now verify that there exist two analogous solutions setting
$\beta=kx^2$ or $\beta=kx^{-1/3}$, where $k$ is an arbitrary real
constant, which we set equal to 2 for convenience. In the first case
the related BPS spacetime and Maxwell field are
\begin{eqnarray}
\dd s^2 &=& -\frac{z^4}{\ell^2(z^2+x^4)} \dd t^2 + \frac{\ell^2(z^2+x^4)}{z^4} \dd z^2
         + \frac{\ell^2(z^2+x^4)}{2x^6} \dd x^2 + \nonumber \\
&& + \frac{2z^2(z^2-6x^4)}{3x^3(z^2+x^4)} \dd t\,\dd y + \frac{7\ell^2(z^4+6z^2x^4-9x^8)}
     {18x^6(z^2+x^4)} \dd y^2\,, \\
{\cal F} &=& \frac{1}{(z^2+x^4)^2}\left[-\frac{2}{\ell}xz^2
             (z^2-x^4) \dd t \wedge \dd x - \frac{2}{\ell} x^6 z\,\dd t \wedge \dd z\; + \right. \nonumber \\
&& \left. + \frac{14}{3}\ell x^3 z\,\dd y \wedge \dd z - \frac{7\ell}{6x^2}(z^2-x^4)(z^2-3x^4)
   \dd x \wedge \dd y \right]\,, \nonumber
\end{eqnarray}
and in the second case we have
\begin{eqnarray}
\dd s^2 &=& -\frac{z^4x^{2/3}}{\ell^2(1+z^2x^{2/3})} \dd t^2 + \frac{2z^2}{3x^{2/3}} \dd t\,
         \dd y + \frac{\ell^2(1+z^2x^{2/3})}{z^4x^{2/3}} \dd z^2 \nonumber \\
&& + \frac{\ell^2(1+z^2x^{2/3})}{9x^2} \dd x^2\,, \\
{\cal F} &=& -\frac{2zx^{1/3}}{\ell(1+z^2x^{2/3})^2} \dd t \wedge \dd z -
             \frac{z^2(1-z^2x^{2/3})}{3\ell x^{2/3}(1+z^2x^{2/3})^2} \dd t\wedge \dd x\,. \nonumber
\end{eqnarray}
A calculation of the Weyl scalars shows that the two spacetimes, as before,
are of Petrov type I and Petrov type N respectively.
We finally stress the fact that these solutions can also be obtained from the
``purely electric'' solutions of section~\ref{electric} by an
appropriate PSL$(2,\bR)$ transformation.

%____________________________________________________________________________%
\subsubsection{Kink solutions and generalizations}

More general solutions can be obtained in the $\beta=\beta(x,y)$ case.
As we mentioned above, the system (\ref{gamma})-(\ref{beta})
follows from the dimensionally reduced gravitational Chern-Simons
action, which (for $k>0$) admits the ``kink'' solution
\cite{Guralnik:2003we}
\begin{equation}
\gamma = -2\ln\cosh\frac{\sqrt k}2 X\,, \qquad \beta = \sqrt k \tanh\frac{\sqrt k}2 X\,,
\end{equation}
where the coordinate $X$ is related to $x$ by
\begin{equation}
x = \frac 1{\sqrt k}\sinh\frac{\sqrt k}2 X \cosh\frac{\sqrt k}2 X + \frac X2\,.
\end{equation}
Let us assume $a\neq 0$ in (\ref{A}) and shift $z \to z - b/2a$. This yields
\begin{equation}
F = \frac{2(z + in\tanh\frac{\sqrt k}2 X)}{z^2 + n^2}\,,
\end{equation}
where we defined $n = \sqrt k/2a$. One can now solve (\ref{omega}) to determine
$\omega_i$. Finally, by rescaling $X \to X/\sqrt k$, $y \to y/\sqrt k$,
$z \to 2nz/\ell$, $t \to \ell t/2n$, we can effectively set $k=1$, $n = \ell/2$
in the supergravity solution, which reads
\begin{eqnarray}
\dd s^2 &=& -\frac 1{\ell^2}\frac{(z^2 + \frac{\ell^2}4)^2}{z^2 + \frac{\ell^2}4
         \tanh^2\frac X2}\left[\dd t + \left(\frac{\ell^3}{4(z^2 + \frac{\ell^2}4)
         \cosh^4\frac X2} - \frac{\ell}{\cosh^2\frac X2}\right)\dd y\right]^2 \nonumber \\
     & & + \ell^2\frac{z^2 + \frac{\ell^2}4\tanh^2\frac X2}{(z^2 + \frac{\ell^2}4)^2}\dd z^2
         + (z^2 + \frac{\ell^2}4\tanh^2\frac X2)(\dd X^2 + \frac{\dd y^2}{\cosh^4\frac X2})\,,
         \label{kink} \\
{\cal A} &=& \frac 12 \frac{z^2 + \frac{\ell^2}4}{z^2 + \frac{\ell^2}4\tanh^2\frac X2}
             \tanh\frac X2\,\dd t\,. \nonumber
\end{eqnarray}
Asymptotically for $X \to \pm \infty$ the gauge field goes to zero and the
metric approaches
\begin{equation}
\dd s^2 \to -\left(\frac{z^2}{\ell^2} + \frac 14\right)\left[\dd t \mp \frac{\ell}u \dd y\right]^2
         + \frac{\dd z^2}{\frac{z^2}{\ell^2} + \frac 14} + \ell^2\left(\frac{z^2}{\ell^2}
         + \frac 14\right)\frac{\dd u^2 + \dd y^2}{u^2}\,, \label{AdSOrtin}
\end{equation}
where we defined the new coordinate $u=\pm e^{\pm X}/4$.
Eqn.~(\ref{AdSOrtin}) is simply AdS$_4$ written in nonstandard
coordinates \cite{Alonso-Alberca:2000cs}, so that the ``kink''
solution (\ref{kink}) interpolates between two AdS vacua at
$X=\pm\infty$.

Grumiller and Kummer were able to write down the most general solution
of (\ref{CS}), using the fact that the dimensionally reduced
gravitational Chern-Simons theory can be written as a Poisson-sigma
model with four-dimensional target space and degenerate Poisson tensor
of rank two \cite{Grumiller:2003ad}. This solution is given by
\cite{Grumiller:2003ad}
\begin{eqnarray}
\gamma &=& -2\ln\cosh\frac{\sqrt k}2 X + \frac 12\ln(1 + \delta)\,, \nonumber \\
\delta &=& (8{\cal C}/k^2 - 1)\cosh^4\frac{\sqrt k}2 X\,, \nonumber \\
\beta &=& \sqrt k \tanh\frac{\sqrt k}2 X\,, \label{grum}
\end{eqnarray}
where the coordinate $X$ is related to $x$ by
\begin{displaymath}
\frac{dx}{dX} = \frac{\cosh^2\frac{\sqrt k}2 X}{1 + \delta}\,,
\end{displaymath}
and ${\cal C}$ denotes an integration constant\footnote{More precisely, ${\cal C}$ and
$k$ are the Casimir functions of the Poisson sigma model that can be interpreted
respectively as energy and charge \cite{Grumiller:2003ad}.}.
In the special case $8{\cal C} = k^2$ we recover the kink solution considered above.
As before, one can now determine $F$ and $\omega_i$ corresponding to (\ref{grum}).
This gives rise to new BPS supergravity solutions generalizing (\ref{kink}).

%%%%%%%%%%%%%%%%%%%%%%%%%SOLUZIONE GENERALIZZATA %%%%%%%%%%%%%%%%%%%%%%
Using the same coordinates as in (\ref{kink}) we find
\begin{eqnarray}
\dd s^2 &=& -\frac 1{\ell^2}\frac{(z^2 + \frac{\ell^2}4)^2}{z^2 + \frac{\ell^2}4
         \tanh^2\frac X2}\left[\dd t + \left(\frac{\ell^3(1+\delta)}{4(z^2 + \frac{\ell^2}4)
         \cosh^4\frac X2} - \frac{\ell}{\cosh^2\frac X2}\right)\dd y\right]^2 \nonumber \\
     & & + \ell^2\frac{z^2 + \frac{\ell^2}4\tanh^2\frac X2}{(z^2 + \frac{\ell^2}4)^2} \dd z^2
         +\frac {(z^2 + \frac{\ell^2}4\tanh^2\frac X2)}{1+\delta}\left(\dd X^2 + \frac{(1+\delta)^2}
         {\cosh^4\frac X2} \dd y^2 \right)\,,
         \nonumber \\
{\cal A} &=& \frac 12 \frac{z^2 + \frac{\ell^2}4}{z^2 + \frac{\ell^2}4\tanh^2\frac X2}
             \tanh\frac X2\,\dd t +\delta_0 \frac {\ell^3}8
             \frac {\tanh\frac X2}{z^2+\frac {\ell^2}4 \tanh^2\frac X2} \,\dd y \,. \label{kink1}
\end{eqnarray}
with $\delta_0 = \frac {8{\cal C}}{k^2} -1$.

If $\delta_0 \leq 0$ then the metric is well-defined only in the region
$-\bar X \leq X \leq \bar X$ where $\cosh({\bar X}/2) = -1/\delta_0$
and $\delta_0 \geq -1$.\\
If $\delta_0 \geq 0$ we can take $X \to \pm \infty$ so that the metric becomes
\begin{displaymath}
\dd s^2 = -\left(\frac{z^2}{\ell^2} + \frac 14\right)\left[\dd t \mp \frac{\ell}u \dd y
+\frac {l^3 \delta_0}{4z^2 +\ell^2} \dd y \right]^2
         + \frac{\dd z^2}{\frac{z^2}{\ell^2} + \frac 14} +
         \ell^2\left(\frac{z^2}{\ell^2}
         + \frac 14\right)\left[ \frac {\dd u^2}{\delta_0 u^4} +\delta_0\dd y^2 \right]\,,
\end{displaymath}
where again $u = \pm e^{\pm X}/4$,
and the gauge field asymptotes to
\begin{equation}
{\cal A} =\frac {\delta_0 \ell}8 \frac {\dd y}{\frac{z^2}{\ell^2} + \frac 14}.
\end{equation}

Note that all these solutions are defined only for $k\geq 0$. One
can now verify that the domain of the parameter $k$ can be
extended also to the negative region. Setting $\sqrt k=i\eta$, in
general one has the following functions
\begin{eqnarray}
\gamma &=& -2\ln\cos\frac{\eta}2 X + \frac 12\ln(1 + \delta)\,, \nonumber \\
\delta &=& (8{\cal C}/\eta^4 - 1)\cos^4\frac{\eta}2 X\,, \nonumber \\
\beta &=& -\eta\tan\frac{\eta}2 X\,. \label{grumkneg}
\end{eqnarray}
This yields
\begin{equation}
F = \frac{2(z - in\tan\frac{\eta}2 X)}{z^2 - n^2}\,,
\end{equation}
where we defined $n\equiv\eta/2a$. As before we can set, after a
diffeomorphism, $k=-1$ and $n=\ell/2$, so that the solution reads
\begin{eqnarray}
\dd s^2 &=& -\frac 1{\ell^2}\frac{(z^2 - \frac{\ell^2}4)^2}{z^2 +
\frac{\ell^2}4
         \tan^2\frac X2}\left[\dd t - \left(\frac{\ell^3(1+\delta)}{4(z^2 - \frac{\ell^2}4)
         \cos^4\frac X2} + \frac{\ell}{\cos^2\frac X2}\right)\dd y\right]^2 \nonumber \\
     & & + \ell^2\frac{z^2 + \frac{\ell^2}4\tan^2\frac X2}{(z^2 - \frac{\ell^2}4)^2} \dd z^2
         +\frac {(z^2 + \frac{\ell^2}4\tan^2\frac X2)}{1+\delta}\left(\dd X^2 + \frac{(1+\delta)^2}
         {\cos^4\frac X2} \dd y^2 \right)\,,
         \label{kink1-neg} \\
{\cal A} &=& -\frac 12 \frac{z^2 - \frac{\ell^2}4}{z^2 +
\frac{\ell^2}4\tan^2\frac X2}
             \tan\frac X2\,\dd t +\delta_0 \frac {\ell^3}8
             \frac {\tan\frac X2}{z^2+\frac {\ell^2}4 \tan^2\frac X2} \,\dd y \,. \nonumber
\end{eqnarray}

%____________________________________________________________________________%
\subsubsection{The case $\beta=\beta(z)$}
\label{betazeta}

In this case $B(x,y,z)$ must be a function of the coordinate $z$
alone, and the system of equations which describes this set of
solutions is
\begin{eqnarray}
B''+3\left(AB\right)'+B\left(3A^2-B^2\right)&=&0,\nonumber\\
\check\RR(\gamma)&=&k,\\
e^{2\int\dd zA(z)}\left[2A'+A^2-3B^2\right]&=&k,\nonumber
\end{eqnarray}
In particular, we have that the two-manifold with metric $\dd
s^2=e^{2\gamma}\left(\dd x^2+\dd y^2\right)$ has a constant scalar
curvature. The general solution with $F=F(z)$ and
$\check\RR(\gamma)$ was studied in \cite{Caldarelli:2003pb}: as
one can verify, the complex variable $F$ has the following form
\begin{equation}
    F=\frac{2az+b}{az^2+bz+c}
\end{equation}
where $a$, $b$ and $c$ are complex integration constants (if these
constants are real, we fall back in the ``purely magnetic'' case
already considered in section~\ref{magnetic}).
In \cite{Caldarelli:2003pb} it was shown that setting $a\neq0$ one
recovers the supersymmetric Reissner-Nordstr\"om-Taub-NUT-AdS$_4$
(RNTN-AdS$_4$) solutions obtained in \cite{Alonso-Alberca:2000cs} (see
eqn.~(\ref{RNTN}) for the relation with the NUT parameter and electric
and magnetic charges). Let's consider now the case $a=0$. We have
\begin{equation}
    F=\frac{b}{bz+c}
\end{equation}
with $b\neq0$. Now we shift
\begin{equation}
    z\rightarrow z-\frac{1}{2}\left(\frac{\bar c}{\bar
    b}+\frac{c}{b}\right)
\end{equation}
and define the real constant
\begin{equation}
    n\equiv\frac{i}{2}\left(\frac{\bar c}{\bar
    b}-\frac{c}{b}\right)
\end{equation}
This yields
\begin{equation}
    A=\frac{z}{z^2+n^2},\qquad B=\frac{n}{z^2+n^2}
\end{equation}
and, as a consequence, $k=-1$ and staticity (i.e. $\dd\omega=0$)
for the four-dimensional solution. The solution reads
\begin{eqnarray}
    \dd s^2&=&-\frac{4(z^2+n^2)}{\ell^2}\dd t^2+\frac{\ell^2}{4(z^2+
    n^2)}\dd z^2+\frac{\ell^2}{2x^2}\left(\dd x^2+\dd y^2\right)\\
    \cal F&=&-\frac{\ell}{2x^2}\dd x\wedge\dd y
\end{eqnarray}
which is AdS$_2\times\HH^2$ with magnetic flux on $\HH^2$, where
AdS$_2$ is written in global coordinates.

%%%%%%%%%%%%%%%% HARMONIC SOLUTIONS %%%%%%%%%%%%%%%%%%%%%%%%%%%%%%%%%%%%
\subsection{Harmonic solutions}
Another rich class of solutions can be found if we choose $F$ to be
harmonic, $\Delta F = 0$. If we admit the possibility of isolated
singularities to be present in the closure of the $(x,y)$ domain,
then this does not require $(x,y)$-independence. Eqn.~(\ref{F})
gives
\begin{equation}
F =\frac {2a z +b}{az^2 +bz +c}\,,
\label{Fharm}\end{equation}
where $a$, $b$ and $c$ are complex functions of $(x,y)$. Introducing
the complex variable $\zeta =x+iy$, the harmonicity condition is
equivalent to require that $a$, $b$ and $c$ are all holomorphic
(or antiholomorphic) functions of $\zeta$.

Next, from (\ref{ReF}) one obtains
\begin{equation}
\phi =\ln |az^2 +bz +c| +\gamma (x,y)\,,
\end{equation}
and (\ref{phi}) becomes
\begin{equation}
\check\RR\left(\gamma\right)=2\left(a\bar c+\bar ac\right) -b\bar b\,,
\label{gammaharm}\end{equation}
showing that the scalar curvature
$\check\RR\left(\gamma\right)=-e^{-2\gamma}\Delta2\gamma$ of the
two-dimensional metric $e^{2\gamma} (\dd x^2 +\dd y^2 )$ is not
constant in general.

For the shift vector we find
\begin{eqnarray}
\omega_x -i\omega_y &=&w_x -iw_y+
\frac {\ell^2}2\int  \partial_\zeta \left(F\bar F\right) \dd z\,,  \\
\partial_z (w_x -iw_y ) &=& 0\,, \\
\dd w &=& -i\frac {\ell^2}2 e^{2\gamma} \left(a\bar b -b\bar
a\right)\dd x\wedge\dd y\,. \label{wu}
\end{eqnarray}
Here, $\dd = \dd x^i \partial_i$ denotes the exterior derivative
in two dimensions.
It follows that the form $w=w_i\dd x^i$ can be expressed as
$w=\check\dd\psi$, where $\check\dd\equiv\dd x^i\epsilon_{ij}\partial_j$, and
$\psi (x,y)$ is a function satisfying
\begin{eqnarray}
\Delta \psi = i \frac {\ell^2}2 e^{2\gamma} (a\bar b -b\bar a)\,.
\label{psi}\end{eqnarray}
Note that, if we add a harmonic function $\psi_0$ to $\psi$,
eqn.~(\ref{psi}) is still solved and the shift vector is just
translated by $w_0$, where $w_0=\check\dd\psi_0$ is a closed form,
$\dd w_0=0$. Therefore, at least locally, $w_0=\dd\upsilon$ for some
function $\upsilon$ and the effect of $\psi_0$ can be reabsorbed in a
diffeomorphism $u\mapsto u+\upsilon$. In other words, only solutions
$\psi$ of eqn.~(\ref{psi}) belonging to distinct cohomology classes
produce physically different solutions.

Let us define the Eddington-Finkelstein-like coordinate
$u=t+\frac {\ell^2}4 \int F\bar F \dd z$. Then, the metric takes the form
\begin{equation}
  \dd s^2 = -\frac4{\ell^2F\bar F}\left(\dd u+w\right)^2
  +2\left(\dd u+w\right)\dd z
  +\frac {\ell^2}4 |2az+b|^2 e^{2\gamma} \left(\dd x^2 +\dd y^2
  \right),
\end{equation}
and for the electromagnetic field one finds
\begin{equation}
{\cal F} =-\frac i{\ell} (\dd u+w)\wedge \dd\left( \frac 1F -\frac
1{\bar F} \right) -\frac {\ell}4 |2az +b|^2 \partial_z \left(
\frac 1F +\frac 1{\bar F} -z \right) e^{2\gamma} \dd x \wedge \dd y\,,
\end{equation}
which has the following potential \eqn {\cal A} =\frac i{\ell}
\left( \frac 1F -\frac 1{\bar F} \right) (\dd u+w) +\frac {\ell}2
\check\dd\gamma \ . \feqn

Let us consider now some particular solutions. If the functions $a$,
$b$ and $c$ are constant, we fall in the cases already studied in
sections~\ref{magnetic} and \ref{betazeta}. More precisely, if $a=0$
we obtain the anti-Nariai spacetime AdS$_2\times\HH^2$, while for
$a\neq0$ the BPS limits of the RNTN-AdS$_4$ family of solutions are
recovered. We will analyse now these two cases allowing for
non-constant functions.

%___________________________________________________________________________%
\subsubsection{Supersymmetric Kundt solutions ($a=0$)}

When $a=0$, eqn.~(\ref{psi}) tells us that $\psi$ is an harmonic
function and therefore, performing a diffeomorphism, we can take
$w=0$. Moreover, eqn.~(\ref{Fharm}) reads
\begin{equation}
    F=\frac b{bz+c}.
\end{equation}
Without loss of generality we can set $b=1$ by rescaling accordingly
the curvature of the transverse two-metric, and eqn.~(\ref{gammaharm})
becomes $\check\RR(\gamma)=-1$. Hence, the transverse two-metric
has a constant negative curvature, and describes (at least locally) an
hyperbolic plane $\HH^2$. We can choose for example the solution
$e^{2\gamma}=2/x^2$. Finally, the metric and gauge field read
\begin{eqnarray}
    \dd s^2&=&-\frac{4}{\ell^2}|z+c|^2\dd u^2 +2\dd u\dd
    z+\frac{\ell^2}{2x^2}\left(\dd x^2+\dd y^2\right)\,,\\
    {\cal A}&=&\frac{i}{\ell}\left(c-\bar c\right)\dd
    u+\frac{\ell}{2x}\dd y\,,
\end{eqnarray}
where $c$ is an arbitrary holomorphic function $c=c(\zeta)$.
This metric is precisely of the Kundt form, describing manifolds
admitting a non-expanding, non-twisting null congruence of geodesics,
and the solution has the interpretation of {\em supersymmetric
  electromagnetic and gravitational waves propagating on anti-Nariai
  spacetime}. This is a particular case of the more general solution
found in \cite{Podolsky:2002sy}. Its Petrov type is II.

%___________________________________________________________________________%
\subsubsection{Supersymmetric Robinson-Trautman solutions ($a\neq0$)}

If, instead, $a\neq0$, we can rewrite the function $F$ in the
following way
\begin{equation}
    F=2\frac{z+\beta}{\left(z+\beta\right)^2-\delta}\,,
\end{equation}
where $\beta\equiv b/2a$ and $\delta\equiv(b^2-4ac)/4a^2$ are two
arbitrary holomorphic functions in $\zeta$.
The system of equations describing this class of supersymmetric
configurations is
\begin{eqnarray}
    \phi&=&\ln\left|(z+\beta)^2-\delta\right|+\gamma(x,y)\,,\\
    \check\RR(\gamma) &=& -4\left[2(\im\,\beta)^2+\re\,\delta\right]\,,\\
    \Delta\psi &=& 2\ell e^{2\gamma}\im\,\beta\,.\label{rtpsi}
\end{eqnarray}
In this case, the solution reads
\begin{equation}
    \dd s^2=-\frac{\left|(z+\beta)^{2}-\delta\right|^2}{\ell^2|z+\beta|^{2}}\left(\dd u+w\right)^{2} 
    +2\left(\dd u+w\right)\dd  z+\ell^{2}|z+\beta|^{2}e^{2\gamma}\left(\dd x^2+\dd y^2\right)\,,\\
\end{equation}
with the following electromagnetic potential
\begin{equation}
    {\cal A}=-\frac{\im\left[\left((z+\beta)^{2}-\delta\right)(z+\bar{\beta})\right]}
    {\ell\left|z+\beta\right|^{2}}\left(\dd u+w\right)+\frac{\ell}{2}\check\dd\gamma\,.
\label{rtgen}\end{equation}
If $\beta$ and $\delta$ are constant functions, then, as already
stressed, this solution is of Petrov type D and belongs to the
RNTN-AdS$_4$ family of solutions. Allowing for a $\zeta$-dependence in
these functions deforms the metric, which acquire a non-vanishing
Weyl scalar $\Psi_4$, signaling the presence of gravitational
radiation. In general, solution (\ref{rtgen}) describes
electromagnetic and gravitational expanding waves propagating on a
supersymmetric RNTN-AdS$_4$ background. To our knowledge, these field
configurations where not known previously in the literature.

We shall work out in the following the simplest of these solutions,
leaving the general analysis for further investigations. Suppose that
$\im\,\beta=0$ (this condition corresponds to put the NUT parameter
$n$ of the RNTN-AdS$_4$ solution to zero in the case of constant $a$,
$b$ and $c$). Since $\beta$ is holomorphic, this implies that
$\beta$ should be a real constant, hereafter named $\kappa$.
It follows from eqn.~(\ref{rtpsi}) that $\psi$ is harmonic and
therefore we can take $w=0$. The resulting spacetime and gauge field are
\begin{eqnarray}
    \dd s^2&=&
    \displaystyle
    -\frac{\left|(z+\kappa)^{2}-\delta\right|^2}{\ell^2(z+\kappa)^{2}}
    \dd u^{2} 
    +2\dd u\dd  z+\ell^{2}(z+\kappa)^{2}e^{2\gamma}
    \left(\dd x^2+\dd y^2\right)\,,\\
    {\cal A}&=&
    \displaystyle\frac{\im\,\delta}{\ell\left(z+\kappa\right)}
    \dd u+\frac{\ell}{2}\check\dd\gamma\nonumber\,,
\end{eqnarray}
with $\check\RR(\gamma)=-4\,\re\,\delta$. The constant $\kappa$ can then be
reabsorbed by the shift $z\rightarrow z-\kappa$ and the solution becomes
\begin{eqnarray}
    \dd s^2&=&
    \displaystyle -\left|\frac z\ell-\frac\delta{\ell z}\right|^2\dd u^{2} 
    +2\dd u\dd  z+\ell^{2}z^{2}e^{2\gamma}\left(\dd x^2+\dd
    y^2\right)\,,    \label{rt1}\\
    {\cal A}&=&\frac{\im\,\delta}{\ell z}\,
    \dd u+\frac{\ell}{2}\,\check\dd\gamma\,.\label{rt2}
\end{eqnarray}
This metric is clearly of the Robinson-Trautman form, describing
manifolds admitting an expanding, non-twisting null congruence of
geodesics. To put it in a more familiar shape, we can define the
function $P=\sqrt{2}\ell^{-1}e^{-\gamma}$ and the operator
$\Delta^{*}\equiv\frac 12P^{2}\Delta$, so that
$\check\RR(\gamma)=2\ell^{2}\Delta^{*}\ln P$ and the solution reads
\begin{eqnarray}
    \dd s^2=-\left[\frac{z^{2}}{\ell^{2}}+\Delta^{*}\ln P+\frac{|\delta|^{2}}
    {\ell^2z^{2}}\right]\dd u^{2} 
    +2\dd u\dd  z+\frac{2z^{2}}{P^{2}}\left(\dd x^2+\dd y^2\right)\,,\\
    {\cal A}=\frac{\im\,\delta}{\ell z}\,
    \dd u-\frac{\ell}{2}\check\dd\ln P\,.
    \qquad\qquad\qquad\qquad\qquad\qquad\qquad\qquad\quad\,
\end{eqnarray}
with $P(x,y)$ any solution of $\Delta^{*}\Delta^{*}\ln P=0$, while
$\im\,\delta$ is determined by the fact that $\delta$ is holomorphic
and its real part is fixed by $\check\RR(\ga)$. This solution
is of Petrov type II and
generalizes the massless and purely gravitational Robinson-Trautman
class of solutions found in \cite{Podolsky:2002sy}, by adding
electromagnetic waves on it. In conclusion, we can interpret the
solution (\ref{rt1}), (\ref{rt2}) as {\em supersymmetric
  electromagnetic and gravitational  expanding waves propagating on
  the BPS Reissner-Nordstr\"om-AdS$_4$ spacetimes}.

%%%%%%%%%%%%%%%%%%%%%%%%%%%%%%%%%%%%%%%%%%%%%%%%%%%%%%%%%%%%%%%%%%%%%%%%%%%%
\section{The lightlike case}

\label{lightlike}

In \cite{Caldarelli:2003pb} it was shown that the general
supersymmetric solution in the lightlike case is an electrovac
travelling wave with metric\footnote{For nonabelian generalizations
see \cite{Cariglia:2003ug}.}
\begin{equation}
  \dd s^2=\frac{\ell^2}{x^2}\left[{\mathcal G}(x,y,u)\,\dd u^2+2\,\dd u\dd v
  +\dd x^2+\dd y^2\right]\,,
\label{lobwave}\end{equation}
and the null electromagnetic field is given by
\begin{equation}
  {\mathcal F}=\dd{\mathcal A} =\varphi'(u)\,\dd u\wedge\dd x\,,\qquad
  {\mathcal A}=\varphi(u)\,\dd x\,.
\end{equation}
Here, the arbitrary function $\varphi'(u)$ defines the profile of the
electromagnetic wave propagating on this metric, while $\G(x,y,u)$ is
any solution of the inhomogeneous Siklos equation \cite{Siklos:1985}
\begin{equation}
  \Delta{\mathcal G}-\frac2x\partial_x{\mathcal G}=-\frac{4x^2}{\ell^2}
  \left(\varphi'\right)^2\,.
\label{siklos}\end{equation}
The dependence of $\G(x,y,u)$ on $u$ describes the profile of the
gravitational wave.
The general solution to this equation was obtained in \cite{Siklos:1985},
and reads (cf.~also appendix \ref{appsiklos})
\begin{equation}
  \G(\zeta,\bar\zeta,u)=\frac12\left(\zeta+\bar\zeta\right)\left(
    \partial f+\bar\partial\bar f\right)-\left(f+\bar f\right)
  -\frac{\left(\varphi'(u)\right)^2}{16\ell^2}\left(\zeta+\bar\zeta\right)^4\,,
  \label{gensol}
\end{equation}
where $f(\zeta,u)$ is an arbitrary holomorphic function in $\zeta=x+iy$.

This family of travelling waves enjoys a large group of coordinate
transformations which preserve the form (\ref{lobwave}) of the line
element. Under the diffeomorphism
$(u,v,x,y)\mapsto(\bar u,\bar v,\bar x,\bar y)$ defined by
\begin{eqnarray}
  \bar u =\chi(u)\,,\qquad \bar x=x\sqrt{\chi'(u)}\,,\qquad
  \bar y=y\sqrt{\chi'(u)}-\psi(u) \,, \phantom{\frac12}\nonumber\\
  \bar v=v-\frac{\chi''(u)}{4\chi'(u)}\left(x^2+y^2\right)
  +\frac{\psi'(u)}{\sqrt{\chi'(u)}}y+\gamma(u)\,,\!\!\!\qquad\qquad\qquad
\label{diff}\end{eqnarray}
where $\chi(u)$, $\psi(u)$ and $\gamma(u)$ are arbitrary functions of
$u$, the metric keeps the same form (\ref{lobwave}) but with
$\bar\G(\bar x,\bar y,\bar u)$ given by\footnote{This invariance was
  found by Siklos \cite{Siklos:1985}. Here we use
  $\chi'(u)=e^{-2\phi(u)}$, and (\ref{Gtrans}) corrects a
  typeset error in equation (34) of \cite{Siklos:1985}.}
\begin{eqnarray}
  \bar\G(\bar x,\bar y,\bar u)=\frac1{\chi'(u)}\left[
    \G(x,y,u)+\frac12\left\{\chi(u);u\right\}\left(x^2+y^2\right)-2\gamma'(u)
  \right]\nonumber\\
  -\frac{2y}{\sqrt{\chi'(u)}}\left(\frac{\psi'(u)}{\chi'(u)}\right)'
  -\left(\frac{\psi'(u)}{\chi'(u)}\right)^2\,.\qquad\qquad\qquad
\label{Gtrans}\end{eqnarray}
Here, the prime denotes the derivative with respect to $u$, while
\begin{equation}
\left\{\chi(u);u\right\}=\frac{\chi'''(u)}{\chi'(u)}-\frac32\left(\frac{\chi''(u)}{\chi'(u)}\right)^2
\end{equation}
defines the Schwarzian derivative. These are not Killing symmetries,
the metric changes, but solutions are brought into other solutions.
The special diffeomorphisms with $\psi(u)=\gamma(u)=0$ correspond to
reparameterizations of the coordinate $u$; this transformation group
is generated by a central extension of the Virasoro algebra
\cite{Banados:1999tw}.

All metrics (\ref{lobwave}) where shown to preserve at least one quarter
of the supersymmetries in
\cite{Caldarelli:2003pb}. Indeed, it is easy to show that the spinor
\begin{equation}
  \epsilon=\frac14\Ga_-\Ga_+\left(1+\Ga_x\right)
  e^{\frac{ix}{\ell}\varphi(u)}\epsilon_0
  \label{commonspinor}
\end{equation}
solves the Killing spinor equation for all these configurations, and
therefore the generalized Siklos spacetimes (\ref{lobwave}) preserve at least one
supersymmetry. Here, $\frac14\Ga_+\Ga_-(1+\Ga_x)$ is a projection operator
of rank one\footnote{The Dirac matrices $\Gamma_{\pm}$ and $\Gamma_x$ are
defined in appendix \ref{appnotations}.}
and $\Ga_+\Ga_-$ is just the usual chirality projector appearing
in the Killing spinors of supersymmetric pp-waves.
The aim of this section is to solve the Killing
spinor equation for these backgrounds, in order to obtain the exact
fraction of supersymmetry preserved, and thus to get a complete
classification of supersymmetric solutions in the lightlike case.

\subsection{First integrability conditions}

As a first step, we shall solve the (first) integrability conditions
$[\DD_\mu,\DD_\nu]\epsilon=0$. Although these conditions are only
necessary \cite{vanNieuwenhuizen:1983wu}, they impose some algebraic
conditions on the functions $\G$ and $\varphi$, and simplify the task
of solving the Killing spinor equations.

The vanishing of the supercurvature yields two nontrivial constraints,
\begin{equation}
  \varphi'(u)\Ga_-\left(1-\Ga_x\right)\epsilon=0\,,
\end{equation}
and
\begin{equation}
  \frac x{2\ell}\left(
    \frac{2ix}\ell\varphi''(u)+\frac12\Delta_-\G+\G_{,xy}\Ga_y
    \right)\Ga_-\epsilon
    -\frac{3i}\ell\varphi'(u)\left(1-\Ga_x\right)
    \epsilon=0\,,
\end{equation}
with $\Delta_-$ defined by $\Delta_- = \partial^2_x - \partial^2_y$.

Using the representation of the Dirac matrices given in
appendix~\ref{appnotations}, which is well-adapted to the present
problem, these relations become respectively
\begin{equation}
 \varphi'(u) \ep_2 =0\,,
\label{int1}
\end{equation}
and
\begin{equation}
  \left(\begin{array}{cccc}
      0 & \frac x{\sqrt{2}\ell}\left(\frac{2ix}{\ell}\varphi''(u)+\frac12\Delta_-\G\right) & \frac x{\sqrt{2}\ell}\G_{,xy} & 0 \\
      0 & \frac{6i}\ell\varphi'(u) & 0 & 0 \\
      0 & 0 & 0 & 0 \\
      0 & \frac x{\sqrt{2}\ell}\G_{,xy}
      & \frac x{\sqrt{2}\ell}\left(\frac{2ix}{\ell}\varphi''(u)
        -\frac12\Delta_-\G\right)
      & \frac{6i}\ell\varphi'(u)
  \end{array}\right)
  \left(\begin{array}{c}
      \ep_1 \\ \ep_2 \\ \ep_3 \\ \ep_4
  \end{array}\right)=0\,,
\label{int2}
\end{equation}
where $\ep_i$ are the spinor components.
The first relation suggest to study two distinct cases, the
purely gravitational one $\varphi'(u)=0$ and the case where
electromagnetic fields are present.

\subsection{Purely gravitational waves}

This case has been extensively studied in \cite{Siklos:1985}
(see \cite{Brecher:2000pa} for a generalization to higher dimensions).
When $\varphi'(u)=0$, only equation (\ref{int2}) yields new conditions
and the rank of this matrix tells us the number of
Killing spinors eliminated by the integrability conditions.
There are two cases to be analysed. First, if
$\left(\Delta_-\G\right)^2+ 4\left(\G_{,xy}\right)^2=0$ the matrix
vanishes. Otherwise its rank is two.

\subsubsection{The maximally supersymmetric case: $\Delta_-\G=\G_{,xy}=0$}
These conditions are satisfied if and only if the spacetime is AdS
\cite{Siklos:1985}. It follows that the most general way of expressing
AdS in the form (\ref{lobwave}) corresponds to $f(\zeta,u)$ at most quadratic
in $\zeta$, and it can be generated by a coordinate transformation
(\ref{diff}) from the metric with $\G=0$.
In this case the matrix in equation (\ref{int2}) vanishes and no
condition is imposed by (\ref{int1}) and (\ref{int2}). In fact, it is
well-known that AdS is a maximally supersymmetric spacetime
\cite{Breitenlohner:jf} and recently it has been shown that this is the
only maximally supersymmetric solution of the model under
consideration \cite{Caldarelli:2003pb}.
For the sake of completeness, we give here the four independent
Killing spinors, which in our conventions read
\begin{eqnarray}
 \displaystyle
  \epsilon^{(1)}=\left(1,0,0,0\right),\quad &
  \displaystyle
  \qquad\epsilon^{(2)}=\left(\sqrt2\frac{v}{\ell},1,\frac yx,0\right),
  \nonumber\\
 \displaystyle
  \epsilon^{(3)}=\left(0,0,\frac1x,0\right),\quad &
  \;\,\epsilon^{(4)}=\left(-y,0,0,x\right).
\end{eqnarray}

\subsubsection{One quarter BPS Lobatchevski waves: $\Delta_-\G\neq0$
or $\G_{,xy}\neq0$}
In this case $f(\zeta,u)$ is at least of order three in $\zeta$ and
the spacetime describes an exact AdS gravitational wave
\cite{Podolsky:1997ik}.
The matrix (\ref{int2}) has rank two and the configuration is at most
one half BPS. However, while solving the Killing spinor equation, an
additional condition on the Killing spinor emerges, leaving just one
quarter of the supersymmetries, as shown in \cite{Brecher:2000pa}.
The residual supersymmetry is generated by the Killing spinor
(\ref{commonspinor}). This is a very simple example that shows that the
vanishing of the supercurvature is not a sufficient condition to ensure the
existence of Killing spinors.

\subsection{The electromagnetic case}

Let us now turn on the electromagnetic field and consider the
case where $\varphi'(u)\neq0$.
The first integrability condition (\ref{int1}) imposes now
$\ep_2=0$.
Then, the second integrability condition (\ref{int2}) simplifies to
the following two relations,
\begin{equation}
  \G_{,xy}\ep_3=0\,,\qquad
  \frac x{\sqrt{2}\ell}\left(\frac{2ix}{\ell}\varphi''(u)
    -\frac12\Delta_-\G\right) \ep_3 +
  \frac{6i}\ell\varphi'(u) \ep_4 =0\,.
\end{equation}
The former relation suggests us to examine separately the cases
$\G_{,xy}\neq0$ and $\G_{,xy}=0$.

\subsubsection{Generic one quarter BPS solution: $\varphi'(u)\neq0$ and
  $\G_{,xy}\neq0$}
  In this case, we have $\ep_3=\ep_4=0$ in addition to $\ep_2=0$ and
  the only Killing spinor available is (\ref{commonspinor}). Hence,
  these spacetimes are exactly one quarter supersymmetric.

\subsubsection{One half BPS waves: $\varphi'(u)\neq0$ and $\G_{,xy}=0$}
The most general solution of the Siklos equation restricted by these
conditions is given by
\begin{equation}
  \G(x,y,u)=-\frac1{\ell^2}\varphi'(u)^2x^4+\frac16\xi_3(u)x^3
  +\frac12\xi_2(u)\left(x^2+y^2\right)+\xi_1(u)y+\xi_0(u)\,,
  \label{Ghalf}\end{equation}
where $\xi_i(u)$ are arbitrary real functions of $u$.
Then, $\Delta_-\G=\xi_3(u)x-\frac{12}{\ell^2}\varphi'(u)^2x^2$
and (\ref{int2}) becomes a relation between the components $\ep_3$ and
$\ep_4$ of the Killing spinor,
\begin{equation}
  \frac x{\sqrt2}\left(\frac{2ix}\ell\varphi''(u)
    +\frac{6x^2}{\ell^2}\varphi'^2(u)-\frac
    x2\xi_3(u)\right)\ep_3
  +6i\varphi'(u)\ep_4=0
  \label{cond}\end{equation}
As we have also $\ep_2=0$, in this case there can be at
most two independent components of the Killing spinor.
Let us solve now the Killing spinor equations to see whether the
existence of the second supersymmetry imposes further constraints.
Some lenghty but straightforward algebra shows that these equations
are solved by
\begin{equation}
  \ep_1=\frac1{\sqrt2\ell}\left(A+B(u)-y\kappa'(u)\right)
         e^{\frac{ix}\ell\varphi(u)}\,\qquad
  \ep_2=0\,,
\end{equation}
\begin{equation}
  \ep_3=\frac{\kappa(u)}xe^{\frac{ix}\ell\varphi(u)}\,,\qquad
  \ep_4=\frac x{\sqrt2\ell}\left(\kappa'(u)
    +\frac{ix}{\ell}\kappa(u)\varphi'(u)\right)e^{\frac{ix}\ell\varphi(u)}\,,
\end{equation}
where $A$ is an arbitrary constant while $B(u)$ and $\kappa(u)$ are two
complex functions subject to the conditions
\begin{eqnarray}
  B'(u)+\frac12\xi_1(u)\kappa(u)=0\,,
  \label{eqB}\\
  \kappa''(u)-\frac12\xi_2(u)\kappa(u)=0\,.
  \label{eqkappa}
\end{eqnarray}
Now the relation (\ref{cond}) can be solved for $\kappa(u)$,
\begin{equation}
  \kappa(u)=\kappa_0\left(\varphi'(u)\right)^{-\frac13}\exp\left(
    -\frac{i\ell}{12}\int\frac{\xi_3(u)}{\varphi'(u)}\dd u
  \right)\,,
\end{equation}
with $\kappa_0$ an arbitrary complex constant,
and we are left with the additional consistency condition
(\ref{eqkappa}).
Since the $\xi_i$'s are real, the real and imaginary parts of
the constraint determine the functions $\xi_2(u)$ and $\xi_3(u)$ to be
\begin{eqnarray}
  \xi_2(u)&=&\frac89\left(\frac{\varphi''(u)}{\varphi'(u)}\right)^2-\frac23
  \frac{\varphi'''(u)}{\varphi'(u)}-2\alpha^2\ell^2\left[\varphi'(u)\right]^{4/3}\nonumber\\
  \xi_3(u)&=&\displaystyle12\alpha\left[\varphi'(u)\right]^{5/3}\,.
  \label{xi}\end{eqnarray}
whith $\al$ an arbitrary real integration constant.
In conclusion, if $\G(x,y,u)$ has the form (\ref{Ghalf}), with
$\xi_2(u)$ and $\xi_3(u)$ given by (\ref{xi}), the solution preserves
exacly one half of the supersymmetries, otherwise it is only one
quarter BPS. The family of half BPS solutions is parameterized by two
real functions $\xi_0(u)$, $\xi_1(u)$ and a real number
$\alpha$. Howerver, many of these solutions are related by
diffeomorphism, and are in fact equivalent.

To understand better the nature of these one half supersymmetric solutions,
we can put them in a canonical form by performing a suitable change of
coordinates. In particular, we choose
\begin{eqnarray}
  \chi(u)=\int\left(\varphi'(u)\right)^{\frac23}\dd u\,,\qquad\qquad\quad\;\,\\
  \psi'(u)=\frac12\left(\varphi'(u)\right)^{\frac23}\int\frac{\xi_1(u)}{\left(\varphi'(u)\right)^{1/3}}\dd u\,,\\
  \ga'(u)=\frac12\xi_0(u)-\frac{\left(\psi'(u)\right)^2}{2\left(\varphi'(u)\right)^{2/3}}\,.\qquad\;
\end{eqnarray}
The effect of $\chi(u)$ is to bring the field strength in the form
$\F=\dd\bar u\wedge\dd\bar x$ and to simplify $\xi_2(u)$; the
functions $\psi(u)$ and $\gamma(u)$ eliminate the linear term in $y$
and the $\xi_0(u)$ term in $\G(x,y,u)$. Finally, the general one half
BPS solution is, up to diffeomorphisms, given by
\begin{equation}
  \G_{\al}(x,y,u)=-\frac{x^4}{\ell^2}+2\al
  x^3-{\al^2\ell^2}\left(x^2+y^2\right)\,,\qquad
  \varphi(u)=u\,, \label{Galpha}
\end{equation}
and parameterized by a single real number $\al$. The nonvanishing
components of the corresponding Killing spinors are
\begin{eqnarray}
  \ep_1=\frac A{\sqrt2\ell}\exp\left[\frac{iux}\ell\right]
  +\frac{i\kappa_0\al}{\sqrt2}y
  \exp\left[\frac{iu}\ell\left(x-\al\ell^2\right)\right]\,, \nonumber \\
  \ep_3=\frac{\kappa_0}x\exp\left[\frac{iu}\ell\left(x-\al\ell^2\right)\right]\,, \\
  \ep_4=\frac{i\kappa_0x}{\sqrt2\ell^2}\left(x-\al\ell^2\right)
  \exp\left[\frac{iu}\ell\left(x-\al\ell^2\right)\right]\,. \nonumber
\end{eqnarray}
The Killing spinor (\ref{commonspinor}), common to all lightlike
solutions, is recovered by setting $\kappa_0=0$ and $A=1$, while the
other independent Killing spinor is obtained by taking $\kappa_0=1$
and $A=0$.

It is finally important to distinguish between the $\al=0$ and
$\al\neq0$ solutions. Siklos has classified the spacetimes of the form
(\ref{lobwave}) according to the number of independent Killing
vectors \cite{Siklos:1985}. It follows that if $\al=0$ the spacetime
admits a five-dimensional group of isometries, generated by five
Killing vectors. In fact, this solution is exactly the solution
(\ref{alpha=-1/3}) of the timelike case, and the groups of isometries
agree.

On the other hand, when $\al\neq0$, the canonical form of $\G$ falls
in the $A(x,y)$ class of \cite{Siklos:1985}, meaning that we have just
the two trivial Killing vectors $\p_u$ and $\p_v$.

In  table \ref{table_lightlike} we summarize the complete classification
of supersymmetric solutions of the lightlike case.

\TABLE{
    \begin{tabular}{|c||c|c|}
      \hline
      {\em Lightlike case} &
      \parbox[c]{5.5cm}{
        \centerline{purely gravitational
      solutions$\vphantom{\displaystyle1^{\displaystyle H}}$}
        \centerline{$\varphi'(u)=0\vphantom{\displaystyle\frac{.}{1}}$}
      } &
      \parbox[c]{5cm}{
        \centerline{electrovac spacetimes$\vphantom{\displaystyle1^{\displaystyle H}}$}
        \centerline{$\varphi'(u)\neq0\vphantom{\displaystyle\frac.1}$}
      } \\
      \hline
      \hline
      One quarter BPS     & Lobatchevski wave & $\G\neq\G_\al$       \\
      \hline
      One half BPS        & \textit{none}              & $\G_\al(u,x,y)$ \hspace{0.2cm} (\ref{Galpha}) \\
      \hline
      Three quarters BPS  & \textit{none}              & \textit{none}        \\
      \hline
      Maximally SUSY  & AdS$_4$           & \textit{none}            \\
      \hline
    \end{tabular}
  \caption{Classification of supersymmetric spacetimes in the lightlike
    case. Note that the fraction 3/4 of supersymmetry cannot be preserved.}
  \label{table_lightlike}
}

\subsection{One half BPS lightlike solutions}

We will now analyze more in detail the one half BPS solutions with $\G = \G_\al$,
(\ref{Galpha}). In particular we can compute the norm squared of the Killing vectors
constructed from the Killing spinors, in order to
see if they also belong to the timelike class of the theory. Using the definition
$V_{\mu}=i\bar\epsilon\Gamma_{\mu}\epsilon$ and choosing the $\Gamma$-matrix representation
given in appendix \ref{appnotations}, we obtain for the components of $V$ in the vierbein frame
(see appendix \ref{geometrylight}),
\begin{eqnarray}
V_{+}&=&-\frac{1}{\sqrt 2\ell^{4}}\left[|k_{0}|^{2}x^{2}\left(x-\alpha\ell^{2}\right)^{2}+
        \ell^{2}\left|A+ik_{0}\alpha\ell ye^{-iu\alpha\ell}\right|^2\right]\,,\nonumber\\
V_{-}&=&\frac{\sqrt{2}|k_{0}|^{2}}{x^2}\,,\nonumber\\
V_{x}&=&0\,,\nonumber\\
V_{y}&=&-\frac{\sqrt{2}}{\ell x}\re\left(A\bar k_{0}e^{iu\alpha\ell}\right)\,.\nonumber
\end{eqnarray}
The norm squared of $V$ is given by
\begin{displaymath}
V^2=-\frac{2|k_0|^2}{\ell^4 x^2}\left[|k_0|^2x^2\left(x-\alpha\ell^2\right)^2+\ell^2\left|A+ik_0\alpha\ell
y e^{-iu\alpha\ell}\right|^2\right]+\frac{2}{\ell^2 x^2}\re\left(A\bar k_0e^{iu\alpha\ell}\right)^2\,.
\end{displaymath}
If $k_0=0$ we have $V^2=0$, so in this case the spinor generates the null vector typical of the lightlike
solutions. Instead, if $A=0$ one obtains
\begin{equation}
V^2=-\frac{2|k_0|^4}{\ell^4 x^2}\left[x^2\left(x-\alpha\ell^2\right)^2+\alpha^2\ell^4 y^2\right]=
\frac{2|k_0|^4}{\ell^2 x^2}\G_{\alpha}\,,
\end{equation}
which is negative, so the solution also belongs to the timelike class.
Therefore every one half supersymmetric lightlike solution is also a timelike solution.
In order to write down this geometry in the form (\ref{metric}), we also need the other
tensors $f = \bar{\epsilon}\epsilon$, $g = i\bar{\epsilon}\Gamma_5\epsilon$ and
$A_{\mu} = i\bar{\epsilon}\Gamma_5\Gamma_{\mu}\epsilon$ used in \cite{Caldarelli:2003pb}.
A little bit of algebra yields
\begin{eqnarray*}
f&=&\frac{\sqrt 2 |k_0|^2}{\ell^2}\left(x-\alpha\ell^2\right)\,,\\
g&=&-\sqrt 2|k_0|^2\alpha\frac yx\,,\\
F&=&\frac{2i}{\ell}\frac{1}{f-ig}=i\frac{\sqrt 2\ell}{|k_0|^2}\frac{1}{\left(x-\alpha\ell^2\right)
                                  +i\alpha\ell^2\frac yx}\,,\\
A_{\mu} \dd x^{\mu}&=&-\frac{\sqrt 2|k_0|^2}{\ell}\dd\left[\frac yx\left(x-\alpha\ell^2\right)\right]\,.
\end{eqnarray*}
For simplicity we set $\sqrt 2|k_0|^2/\ell=1$, so that one has $V^2=\G_{\alpha}/x^2$,
and $V=-\ell^{-1}\partial_u$.
Comparing this with $V = \partial_t$ \cite{Caldarelli:2003pb}, we can then identify
$t\equiv-\ell u$ and $\omega\equiv-\ell\G_{\alpha}^{-1}\dd v$. Furthermore,
$A_{\mu}\dd x^{\mu} = \dd z$ \cite{Caldarelli:2003pb} implies
$z=-\frac yx\left(x-\alpha\ell^2\right)$. The diffeomorphism
\begin{equation}
Y = -\frac{z^2\alpha\ell x^2}{2(x - \alpha\ell^2)^2} + \frac{x^3}{3\ell} - \frac{\alpha\ell x^2}2
    + C\,, \qquad X = v\,, \label{diffeo}
\end{equation}
where $C$ denotes an arbitrary constant, brings the metric finally into the form
(\ref{metric}), with coordinates $z, X, Y$, and with $F$, $\phi$ and $\omega$ given by
\begin{equation}
F = \frac{2i(x - \alpha\ell^2)}{(x - \alpha\ell^2)^2 - i\alpha\ell^2 z}\,, \qquad
    e^{2\phi} = \frac{\ell^2}{x^4}\,, \qquad \omega =  \frac{\ell^3 F\bar F}{4 x^2}\dd X\,.
    \label{Fphiomega}
\end{equation}
In (\ref{Fphiomega}), $x$ is to be understood as a function of $Y$ and $z$ defined by
(\ref{diffeo}). We have checked explicitely that $F$, $\phi$ and $\omega$ satisfy
the equations (\ref{F})--(\ref{omega}). The electromagnetic field strength reads
${\cal F} = \ell^{-1}\dd x \wedge \dd t$.

\subsection{Lifting to eleven dimensions}

We can lift the general lightlike geometry (\ref{lobwave}) to a
solution of eleven-dimensional supergravity using the results of
\cite{Chamblin:1999tk,Cvetic:1999xp}. The reduction ansatz for the
metric reads
\begin{equation}
\dd s_{11}^2 = \dd s_4^2 + 4\ell^2 \sum_{i=1}^4\left[\dd\mu_i^2 + \mu_i^2
            \left(\dd\phi_i + \frac{1}{2\ell}{\cal A}\right)^2\right]\,,
\end{equation}
where the round metric on the seven-sphere is written as
\begin{displaymath}
\dd\Omega_7^2 = \sum_{i=1}^4 \left(\dd\mu_i^2 + \mu_i^2 \dd\phi_i^2\right)\,.
\end{displaymath}
The $\mu_i$, which satisfy $\sum_i \mu_i^2 = 1$, can be parametrized in
terms of angles on the three-sphere as
\begin{displaymath}
\mu_1 = \sin\theta\,, \quad \mu_2 = \cos\theta\sin\phi\,, \quad
\mu_3 = \cos\theta\cos\phi\sin\psi\,, \quad \mu_4 = \cos\theta\cos\phi\cos\psi\,.
\end{displaymath}
The reduction ansatz for the four-form field strength is given by
\begin{equation}
{\cal F}_{[4]} = -\frac{3}{\ell}\epsilon_{[4]} - 2\ell^2\sum_{i=1}^4 d(\mu_i^2) \wedge
          d\phi_i \wedge ^{\ast}\!{\cal F}\,,
\end{equation}
with $\ast$ denoting the Hodge dual with respect to the four-dimensional metric
and $\epsilon_{[4]}$ its volume form.

In our case we find it convenient to choose ${\cal A} = -x\varphi'(u)\,\dd u$. The
dual of the electromagnetic field strength reads
\begin{displaymath}
^{\ast}{\cal F} = \varphi'(u) \dd u \wedge \dd y\,.
\end{displaymath}
This yields finally for the eleven-dimensional metric, the four-form field
strength and the three-form gauge potential respectively
\begin{eqnarray}
\dd s_{11}^2 &=& \frac{\ell^2}{x^2}\left[{\mathcal G}(x,y,u)\,\dd u^2 + 2 \dd u \dd v
              + \dd x^2 + \dd y^2\right] \nonumber \\
          & & \qquad + 4\ell^2 \sum_i \left[\dd \mu_i^2 + \mu_i^2
              \left(\dd\phi_i - \frac{1}{2\ell}x\varphi'(u)\dd u\right)^2\right]\,, \label{11d} \\
{\cal F}_{[4]} &=& -\frac{3}{\ell}\epsilon_{[4]} + 2\ell^2\varphi'(u)\sum_i \dd(\mu_i^2) \wedge
            \dd\phi_i \wedge \dd u \wedge \dd y\,, \nonumber \\
\mathcal{A}_{[3]} &=& \frac{\ell^3}{x^3}\dd u \wedge \dd v \wedge \dd y + 2\ell^2\varphi'(u)\sum_i
            \mu_i^2 \dd\phi_i \wedge \dd u \wedge \dd y\,. \nonumber
\end{eqnarray}
A special case appears for ${\mathcal G}(x,y,u) = -x^4/\ell^2$ (which is obtained
for $\alpha = 0$ from the half supersymmetric solutions (\ref{Galpha})).
The metric in eleven dimensions is then given by
\begin{displaymath} 
\dd s_{11}^2 = \frac{\ell^2}{x^2}\left[2\dd u \dd v + \dd x^2 + \dd y^2\right] + 4\ell^2 \dd\Omega_7^2
            - 4\ell x\sum_i \mu_i^2 \dd\phi_i\,\dd u\,,
\end{displaymath}
which differs from AdS$_4$ $\times$ S$^7$ only by the last term, which describes
rotation along the S$^7$.

It would be interesting to see exactly how many of the 32 supercharges are preserved by
the solution (\ref{11d}). (We know that they preserve at least two real supercharges
if ${\mathcal G} \neq {\mathcal G}_{\alpha}$ and four if
${\mathcal G} = {\mathcal G}_{\alpha}$, but there might be more). Furthermore,
the form of (\ref{11d}) suggests that the eleven-dimensional solutions might
have an interpretation as the near-horizon limit
of rotating M2-branes with a gravitational wave along the brane. We will leave
these points for future investigations.

%%%%%%%%%%%%%%%%%%%%%%%%%%%%%%%%%%%%%%%%%%%%%%%%%%%%%%%%%%%%%%%%%%%%%%%%%%%%

\section{Final remarks}

\label{finalrem}

In section \ref{lightlike} a complete classification of
the lightlike case was obtained: we showed that the solutions preserve either
one quarter or one half of the supersymmetry, and that a fraction of
three quarters is not possible. The explicit form of the wave profile
for a geometry with two supercovariantly constant spinors was given.
In this case, it was shown that the second Killing spinor gives rise to
a timelike Killing vector.
It would be nice to have a complete classification of the timelike case as
well. For instance, it might be that BPS solutions with three supercovariantly
constant spinors exist in that subclass. In order to obtain such a
classification, one should plug the general form (\ref{metric})
into the integrability conditions
$[{\cal D}_{\mu}, {\cal D}_{\nu}]\epsilon = 0$, and determine the rank of
these matrices. Unfortunately, the timelike
solution (\ref{metric}) is much less explicit than the lightlike one,
eqn.~(\ref{lobwave}), so technically this task is not so easy.

We saw in section \ref{timelikecase} that a kind of ``dimensional
reduction'' (i.~e.~, $z$-independence) of the general equations
(\ref{F}) -- (\ref{ReF}) yields the dimensionally reduced gravitational
Chern-Simons action. This leads us to ask the question if the complete
system (\ref{F}) -- (\ref{ReF}) is described by the gravitational Chern-Simons
theory in three dimensions \cite{Deser:1981wh}.
It would be interesting to pursue this point further,
since a complete understanding of the mathematical structure behind this
set of differential equations would simplify significantly the explicit
construction of all supersymmetric solutions.

Finally, there is an interesting aspect related to the PSL$(2,\bR)$
transformations (\ref{PSLz}), (\ref{PSLphi}). Some months ago, Witten showed
that there is a natural action of the group SL$(2,\bZ)$ on the space of
three-dimensional conformal field theories with U(1) symmetry and a chosen
coupling to a background gauge field \cite{Witten:2003ya}. He argued that
for CFTs with AdS$_4$ dual, the SL$(2,\bZ)$ action on the three-dimensional
CFT may be viewed as the holographic image of the well-known SL$(2,\bZ)$
duality of electrodynamics on AdS$_4$. More recently, Leigh and Petkou
showed that the group SL$(2,\bZ)$ acts also on the two-point functions of
the energy-momentum tensor of three-dimensional CFTs, and suggested
that the holographic image of this action (if any) should be an
appropriate generalization of electromagnetic duality invariance to
include also gravity. Now, looking at (\ref{transfpar}),
we see that our SL$(2,\bR)$ transformations mix electric and magnetic charges,
but they act also on the gravitational field.
They might thus (in the case of supersymmetry) be a candidate for the mentioned
generalization of electromagnetic duality
at the nonlinear level, which includes also gravity. A possible check
of the relevance of (\ref{PSLz}), (\ref{PSLphi}) in this context
would be to do a Fefferman-Graham expansion of the electromagnetic field
strength ${\cal F}$ near the boundary, and see if the PSL$(2,\bR)$ transformations
exchange the two different possible boundary conditions.

\acknowledgments

This work was partially supported by INFN, MURST and
by the European Commission RTN program
HPRN-CT-2000-00131, in which S.~L.~C.~, M.~M.~C.~and D.~K.~are
associated to the University of Torino. We are grateful to L.~Vanzo
for useful discussions.
\normalsize

\appendix

%%%%%%%%%%%%%%%%%%%%%%%%%%%%%%%%%%%%%%%%%%%%%%%%%%%%%%%%%%%%%%%%%%%%%%%%%%%%%%
\section{Conventions}
\label{appnotations}

Throughout this paper, the conventions are as follows:
$a,b,\ldots$ refer to $D=4$ tangent space indices, and $\mu,\nu,\ldots$
refer to $D=4$ world indices. The signature is $(-,+,+,+)$,
$\eps_{0123}=+1$.

The gamma matrices are defined to satisfy the four-dimensional Clifford algebra
$\{\Ga_a,\Ga_b\}=2\eta_{ab}$, and the parity matrix is $\Ga_5 = i\Ga_{0123}$.
We antisymmetrize with unit weight, i.~e.~$\Ga_{ab} \equiv
\Ga_{\left[a\right.}\Ga_{\left.b\right]} \equiv \frac{1}{2}[\Ga_a,\Ga_b]$ etc.
The Dirac conjugate is defined by $\bar\psi=i\psi^\dagger\Ga^0$.

Late latin indices $i,j,\ldots$ refer to two-dimensional submanifolds.
They can take the values $1$, $2$, and $\eps_{12}=+1$.

In Section~\ref{lightlike}, we use a prime to denote differentiation
with respect to $u$, the complex coordinate $\zeta$ is defined by
$\zeta=x+iy$, and $\p$, $\bar\p$ indicate the derivation with respect
to $\zeta$ and $\bar\zeta$ respectively.
The differential operator $\Delta_-$ is defined by
\begin{equation}
\Delta_-=\p_x^2-\p_y^2=2\left(\p^2+\bar\p^2\right)\,.
\end{equation}
Finally, to write down the matrix form of the Killing spinor
equations we use the following representation of the Dirac
matrices,
\begin{eqnarray}
  \Ga_0=\left(\begin{array}{cc}
      i\si_2 & 0\\
      0 & -i\si_2
    \end{array}
  \right)\,,
  && \qquad
  \Ga_1=\left(\begin{array}{cc}
      \si_3 & 0\\
      0 & \si_3
    \end{array}
  \right)\,,
  \\
  \Ga_2=\left(\begin{array}{cc}
      0 & i\si_2\\
      -i\si_2 & 0
    \end{array}
  \right)\,,
  && \qquad
  \Ga_3=\left(\begin{array}{cc}
      -\si_1 & 0\\
      0 & -\si_1
    \end{array}
  \right)\,,
\end{eqnarray}
with
\begin{equation}
  \Ga_5=\left(\begin{array}{cc}
      0 & \sigma_{2}\\
      \sigma_{2} & 0
    \end{array}
  \right)\,,
\end{equation}
and $\ep_1,\ldots,\ep_4$  are the components of the spinor in this
basis.
%%%%%%%%%%%%%%%%%%%%%%%%%%%%%%%%%%%%%%%%%%%%%%%%%%%%%%%%%%%%%%%%%%%%%%%%%%%%%%
\section{PSL$(2,\bR)$ transformations}
\label{appPSL2R}

In this section we show that the PSL$(2,\bR)$ transformations (\ref{PSLz}),
(\ref{PSLphi}) can be used to generate new solutions from known ones.
As an example we use the supersymmetric Reissner-Nordstr\"om-Taub-NUT-AdS$_4$
solution \cite{Alonso-Alberca:2000cs}, which is obtained for \cite{Caldarelli:2003pb}
\begin{eqnarray}
F &=& \frac{2(z + in)}{(z + in)^2 - i\ell(Q - iP)}\,, \nonumber \\
e^{2\phi} &=& \frac{4[(z + in)^2 - i\ell(Q - iP)][(z - in)^2 + i\ell(Q + iP)]}
              {\ell^2 [1 + (x^2 + y^2)^2]}\,, \label{FphiTN}
\label{RNTN}\end{eqnarray}
where $n$ denotes the NUT parameter, and $Q$ and $P$ are the electric and
magnetic charges respectively. The latter is subject to the charge quantization condition
\begin{displaymath}
-2\ell P = \ell^2 + 4n^2\,,
\end{displaymath}
which comes from eqn.~(\ref{phi}). We see that $e^{2\phi}(dx^2 + dy^2)$ represents
the metric on the round two-sphere, and for simplicity we will consider this
case only, although generalizations to hyperbolic or flat spaces
exist \cite{Alonso-Alberca:2000cs}, \cite{Caldarelli:2003pb}.

After applying a PSL$(2,\bR)$ transformation, $F$ and $\phi$ take again
the form (\ref{FphiTN}), but with different parameters $\tilde n$, $\tilde Q$
and $\tilde P$ that are related to $n, Q, P$ by
\begin{eqnarray}
\tilde n &=& |\lambda|\,{\mathrm{Im}}\,\alpha\,, \nonumber \\
\ell\, \tilde Q &=& |\lambda|^2\,(2{\mathrm{Re}}\,\alpha\,{\mathrm{Im}}\,\alpha -
                  {\mathrm{Im}}\,\beta)\,, \label{transfpar} \\
\ell\,\tilde P &=& |\lambda|^2\,({\mathrm{Re}}^2\alpha - {\mathrm{Im}}^2\alpha
             - {\mathrm{Re}}\beta)\,, \nonumber
\end{eqnarray}
where the complex numbers $\lambda$, $\alpha$, $\beta$ are defined by
\begin{eqnarray}
\lambda &=& (d - inc)^2 - i\ell c^2 (Q - iP)\,, \nonumber \\
\alpha &=& \lambda^{-1}[(d - inc)((-b + ina) + cai\ell(Q - iP)]\,, \nonumber \\
\beta &=& \lambda^{-1}[(-b + ina)^2 - a^2 i\ell(Q - iP)] \nonumber
\end{eqnarray}
respectively. To be precise, in order to obtain again exactly the form
(\ref{FphiTN}) after the PSL$(2,\bR)$ transformation, one also has to shift
$z \to z - {\mathrm{Re}}\,\alpha$ and subsequently rescale $z \to z/|\lambda|$,
$t \to |\lambda|\,t$ in the supergravity solution. It is straightforward to
show that the transformed parameters satisfy again the magnetic charge
quantization condition.

We see that the transformation mixes the parameters in a nonlinear way.
In particular, we can start from a solution with vanishing NUT-parameter
and generate one with nonzero $n$. These two solutions are clearly different
topologically, and therefore in general the PSL$(2,\bR)$ transformations
are not merely diffeomorphisms from the four-dimensional point of view.

Although electromagnetic duality invariance is broken in the gauged theory
due to the minimal coupling of the gravitini to the graviphoton, a generalized
duality invariance was discovered in the supersymmetric subclass of the
Pleba\'{n}ski-Demia\'{n}ski solution, which rotates also the mass parameter
into the NUT charge and vice-versa \cite{Alonso-Alberca:2000cs}.
It would be interesting to see whether this duality is a consequence of the
PSL$(2,\bR)$ invariance of the equations (\ref{F}) -- (\ref{ReF}).

%%%%%%%%%%%%%%%%%%%%%%%%%%%%%%%%%%%%%%%%%%%%%%%%%%%%%%%%%%%%%%%%%%%%%%%%%%%%%%
\section{General solution to the inhomogeneous Siklos equation}
\label{appsiklos}

In this section we briefly review the construction of the general solution to the
inhomogeneous Siklos equation (\ref{siklos}) for the function ${\mathcal G}(x,y,u)$,
obtained in \cite{Siklos:1985}.
This is a linear second order partial differential equation, and has a
particular solution given by
\begin{equation}
  \G_0(x,y,u)=-\frac{\left(\varphi'(u)\right)^2}{16\ell^2}\left(\zeta+\bar\zeta\right)^4\,.
\end{equation}
Therefore, the generic solution reads
\begin{equation}
\G(x,y,u)=\G_0(x,y,u)+H(x,y,u),
\end{equation}
where $H$ is the general solution to the homogeneous problem,
\begin{equation}
  \Delta H-\frac2x\partial_xH=0\,.
\end{equation}
To characterize these solutions, one can define
\begin{equation}
  \tilde H=x\int\frac{H(x,y,u)}x^2\dd x\,,
\end{equation}
and the homogeneous Siklos equation reduces to
\begin{equation}
  x^2\p_x\left(\frac{\Delta\tilde H}{x}\right)=0\,,
\end{equation}
or, in other words, $\tilde H$ has to satisfy the Poisson equation
\begin{equation}
  \Delta\tilde H=xK_{,yy}(y,u)\,,
\label{eqhtilde}\end{equation}
where $K(y,u)$ is an arbitrary function of its arguments.
The simplest example of such a function is harmonic,
i.e. $\tilde H=f(\zeta,u)+\bar f(\bar z,u)$, with $f(\zeta,u)$ an arbitrary
analytic function in $\zeta$, in which case we obtain the class of
solutions
\begin{equation}
  \G(\zeta,\bar\zeta,u)=\frac12\left(\zeta+\bar\zeta\right)\left(
    \partial f+\bar\partial\bar f\right)-\left(f+\bar f\right)
  -\frac{\left(\varphi'(u)\right)^2}{16\ell^2}\left(\zeta+\bar\zeta\right)^4\,.
\end{equation}
A particular solution to (\ref{eqhtilde}) is given by
$\tilde H=xK(y,u)$, and since it is a linear partial differential
equation, it follows that its general solution is given by
$\tilde H=xK(y,u)+f(\zeta,u)+\bar f(\bar z,u)$, with $f(\zeta,u)$ a
general holomorphic function in $\zeta$.

Plugging this general solution back into the definition of
$H$, we see that the $xK$ term vanishes and (\ref{gensol}) is the
general solution to our problem, as previously stated.

%%%%%%%%%%%%%%%%%%%%%%%%%%%%%%%%%%%%%%%%%%%%%%%%%%%%%%%%%%%%%%%%%%%%%%%%%%%%%%
\section{Geometry of the lightlike case}
\label{geometrylight}
For the geometry (\ref{lobwave}), the vierbein can be chosen to be
\begin{equation}
  e^+=\frac{\ell^2}{x^2}\,du\,,\qquad
  e^-=dv+\frac12\G\,du\,,\qquad
  e^x=\frac{\ell}{x}\,dx\,,\qquad
  e^y=\frac{\ell}{x}\,dy\,,
\end{equation}
and the spin connection reads
\begin{equation}
  \begin{array}{l@{\qquad}l@{\qquad}l}
    \displaystyle
    \om_{+-}=-\frac1\ell\,e^x\,,
    &
    \displaystyle
    \om_{-x}=\frac1\ell\,e^+\,,
    &
    \displaystyle
    \om_{xy}=\frac1\ell\,e^y\,,
    \\
    &&
    \\
    \displaystyle
    \om_{-y}=0\,,
    &
    \displaystyle
    \om_{+x}=\frac{x^3}{2\ell^3}\,\G_{,x}\,e^+-\frac1\ell\,e^-\,,
    &
    \displaystyle
    \om_{+x}=\frac{x^3}{2\ell^3}\,\G_{,y}\,e^+\,.
  \end{array}
\end{equation}
Finally, the supercovariant derivative becomes
\begin{eqnarray}
  && \displaystyle
  \DD_u=\p_u+\frac\ell{2x^2}\Ga_+(1-\Ga_x)+\frac1{4\ell}\G\Ga_-
  +\frac1{4\ell}\left(x\G_{,x}-\G\right)\Ga_{-x}+\frac x{4\ell}\G_{,y}\Ga_{-y}
  +\frac{ix}{2\ell}\varphi'(u)\Ga_-\Ga_{x+}\,,
  \nonumber\\
  && \displaystyle
  \DD_v=\p_v+\frac1{2\ell}\Ga_-(1-\Ga_x)\,,
  \nonumber\\
  && \displaystyle
  \DD_x=\p_x-\frac i\ell\varphi(u)+\frac1{2x}\left(\Ga_{+-}+\Ga_x\right)
  +\frac{ix^2}{2\ell^2}\varphi'(u)\Ga_-\,,
  \nonumber\\
  && \displaystyle
  \DD_y=\p_y+\frac1{2x}\Ga_y\left(1-\Ga_x\right)
  +\frac{ix^2}{2\ell^2}\varphi'(u)\Ga_-\Ga_{xy}\,.
\end{eqnarray}

\end{document}